\documentclass[12pt,preprint,preprintnumbers,amsfonts,aps,nofootinbib]{article}

\pdfoutput=1 

\usepackage{amssymb}
\usepackage{amsmath}
\numberwithin{equation}{section}
\usepackage{epsfig}

\usepackage{epsf}
\usepackage{graphicx}
\usepackage{amsfonts}
\usepackage{amssymb}
\usepackage{cite}
\usepackage{float}
\usepackage[colorlinks=true,allcolors=blue]{hyperref}
\hypersetup{
    colorlinks=true,
    linkcolor=blue,
    filecolor=blue,      
    urlcolor=blue,
    pdftitle={Beyond general relativity: gravitational waves in non-minimally coupled theories},
    pdfpagemode=FullScreen,
    }
\usepackage{color}

\def\tV{\tilde{\mathcal{V}}}
\def\bh{\bar{h}}
\def\Ve{\mathcal{V}_e}
\def\Vm{\mathcal{V}_m}
\def\tVe{\tilde{\mathcal{V}}_e}
\def\tVm{\tilde{\mathcal{V}}_m}

\def\tx{{\tilde\xi}}
\def\xo{\xi_1}
\def\xt{\xi_2}
\def\xth{\xi_3}
\def\LE{\Lambda_E}
\def\LO{\Lambda_O}

\makeatletter
\renewcommand\section{\@startsection {section}{1}{\z@}%
                                 {-3.5ex \@plus -1ex \@minus -.2ex}
                                   {2.3ex \@plus.2ex}%
                                   {\normalfont\large\bfseries}}
\renewcommand\subsection{\@startsection{subsection}{2}{\z@}%
                                   {-3.25ex\@plus -1ex \@minus -.2ex}%
                                     {1.5ex \@plus .2ex}%
                                     {\normalfont\bfseries}}
\renewcommand\subsubsection{\@startsection{subsubsection}{3}{\z@}%
                                   {-3.25ex\@plus -1ex \@minus -.2ex}%
                                     {1.5ex \@plus .2ex}%
                                     {\normalfont\itshape}}
\makeatother

\def\beq{\begin{equation}}
\def\eeq{\end{equation}}
\def\be{\begin{equation}}
\def\ee{\end{equation}}
\def\bea{\begin{eqnarray}}
\def\eea{\end{eqnarray}}

\def\tBk{\tilde{B}}

\def\tB{\tilde{\mathcal{B}}}
\newcommand{\TM}[1]{{\color{blue}{TM: #1}}} 
\DeclareRobustCommand{\SkipTocEntry}[4]{}
 
\DeclareRobustCommand{\SkipTocEntry}[5]{}

\begin{document}

\begin{titlepage}

\setcounter{page}{1} \baselineskip=15.5pt \thispagestyle{empty}

\begin{flushright}

\end{flushright}
\vfil

\begin{center}

{\Large \bf Beyond general relativity: gravitational waves in non-minimally coupled theories}
\\[0.9cm]
{Stephon Alexander$^{\dag,1}$, Tatsuya Daniel$^{\ast,1,2}$ and Tucker Manton$^{\star,3,1}$}
\\[0.5 cm]

{\small {\sl  $^1$Brown Center for Theoretical Physics \& Innovation, Department of Physics,  
Brown University, Providence, RI 02912, USA}}\\

\vspace{2mm}
{\small {\sl  $^2$Trottier Space Institute, Department of Physics,  
McGill University, Montreal, QC H3A 2T8, Canada}}\\

\vspace{2mm}

{\small {\sl  $^3$School of Fundamental Physics and Mathematical Sciences, Hangzhou Institute for Advanced Study, University of Chinese Academy of Sciences (UCAS-HIAS), Hangzhou 310024, China}}\\

\vspace{.3cm}

\end{center}

\vspace{.8cm}

\hrule \vspace{0.3cm}
{\small  \noindent \textbf{Abstract} \\[0.3cm]
\noindent Non-minimal couplings between matter and curvature tensors arise in many different contexts. Such couplings modify solutions of general relativity (GR) and therefore can be probed in various astrophysical systems. A particularly interesting scenario arises if dark matter experiences non-minimal couplings, as dark matter densities are expected to spike in the vicinity of binary black hole mergers. This gives a novel setting for simultaneously studying dark matter and (beyond) GR physics via observations of gravitational waves (GWs). In this work, we explore effects of various non-minimal couplings on GWs by working with a model-independent parameterization for left- and right-handed GW strains. We extend the parameterization proposed in \cite{Jenks:2023pmk,Daniel:2024lev} to include early-universe effects, and we write down the generic solution assuming slowly-varying matter fields. We then systematically apply our results to three models: Kalb-Ramond dark matter with dimension-four operators, axion-dilaton-Chern-Simons-Gauss-Bonnet dimension-five operators, and dimension-six couplings to a (dark) vector field. 

}\vspace{0.5cm}  \hrule

\vfil

\begin{flushleft}
{\normalsize { \sl \rm \small{$^\dag$~\href{mailto:stephon_alexander@brown.edu}{stephon\_alexander@brown.edu} \\ $^\ast$~\href{mailto:tdaniel@physics.mcgill.ca}{tdaniel@physics.mcgill.ca} \\ $^\star$~\href{mailto:tucker_manton@ucas.ac.cn}{tucker\_manton@ucas.ac.cn}}}}\\

\end{flushleft}

\end{titlepage}

\newpage

{
\hypersetup{linkcolor=black}
\tableofcontents
}

\section{Introduction}\label{sec:Intro}

The detection of gravitational waves (GWs) from merger processes by the LIGO-Virgo-KAGRA (LVK) collaboration \cite{LIGOScientific:2018mvr,LIGOScientific:2020ibl,KAGRA:2021vkt} marked the beginning of a new era of observational cosmology. The earth-based interferometers observe GWs that are emitted during the merging of black holes and/or neutron stars in the ultra-strong field regime. It is precisely in this regime where predictions from general relativity (GR) may fail and as a result, any deviations from GR need to be probed and ideally, with great precision. Such deviations may include higher curvature corrections, higher derivative corrections, or non-minimal matter couplings\footnote{See e.g. \cite{BarrosoVarela:2024ozs, Verner:2024agh, Silveravalle2025} regarding possible cosmological implications of non-minimal (dark) matter couplings to gravity.}. We will refer to all of these possibilities as \textit{beyond}-GR. 

Although GR has so far been in excellent agreement with solar system tests \cite{Will:2014kxa}, pulsar timing \cite{Stairs:2003eg,Hu:2024wub}, and the physics of mergers from compact objects broadly speaking \cite{LIGOScientific:2021sio}, it also predicts pathological spacetime singularities \cite{Penrose:1964wq} and lacks compatibility with quantum mechanics \cite{Carlip:2001wq}. What, if anything, lies beyond GR has been an open question for arguably a century, and GW observations are finally giving the community essential data with which to probe this question in highly energetic systems.

A related, immense open question concerns the nature of dark matter (DM). The only known feature of DM to date is that it experiences gravitational interactions. Due to stable timelike orbits, it is expected that DM accumulates around compact objects such as black holes and neutron stars \cite{Ireland:2024lye, Eroshenko:2024dtb}. This makes the prospect of indirectly detecting DM using GW observations from merger processes - where we may expect a spike in DM density - particularly intriguing \cite{ Wilcox:2024sqs, Chakravarti:2025xaj,Liu:2021xfb,Liu:2022ygf,Liu:2023oab,Liu:2023vno,Liu:2024bfj,Liu:2024xcd,Liu:2025otw,Yang:2023tip,Suarez-Fontanella:2024epb,Owen:2025odr,Alloqulov:2025ucf,Chase:2025wwj,Chen:2025jch,Feng:2025fkc}. 

Such physics can be probed straightforwardly if DM happens to experience non-minimal couplings to gravity \cite{Jenks:2023pmk,Daniel:2024lev,Manton:2024hyc}, which can be constrained by studying modifications to the standard GW waveforms predicted by GR. For example, by studying how GW propagation is modified for a wide class of models, the parameters of any given theory can be mapped to observational constraints on GWs via a model-agnostic framework first developed in \cite{Jenks:2023pmk} and extended in \cite{Daniel:2024lev}\footnote{See also very recent work \cite{Watarai:2025hsb} focusing on probing nonlinear effects using parameterized waveform modeling.}. Furthermore, non-minimal couplings qualitatively either extract or inject energy from the matter sector into the gravitational system, thereby offering potential insight into other mysteries within GR such as dark energy or the final parsec problem \cite{Milosavljevic:2002ht}\footnote{It was shown \cite{Calza:2025yfm} very recently that non-minimal couplings can alter black hole evaporation history, which is relevant for primordial black holes as DM candidates.}.

Non-minimal curvature couplings to DM have been studying in various contexts in the past few decades. Such concepts were considered early on in the context of Modified Newtonian Dynamics (MOND) in \cite{Bruneton:2008fk}, while in \cite{Bettoni:2011fs}, a thorough analysis was performed regarding the deviations from the standard $\Lambda$CDM predictions for linear couplings to the Ricci scalar and Ricci tensor. Some of the same authors studied structure formation in \cite{Bettoni:2012xv} and dark Bose-Einstein Condensates in \cite{Bettoni:2013zma} within non-minimally coupled theories, the latter analysis extended to GWs in \cite{Ivanov:2019iec}. Non-minimal couplings modify the Poisson equation for the gravitational potential, which leads to interesting behavior for DM halos \cite{Gandolfi:2021jai}, galaxy dynamics \cite{Gandolfi:2022puw}, and DM solitons \cite{Zhang:2024bjo}. In fact, this feature of non-minimally coupled DM can result in DM halos that better fit observed profiles than standard cold DM in certain cases \cite{Gandolfi:2023hwx}.

A key goal of the current work is to generalize the parameterizations given in \cite{Jenks:2023pmk,Daniel:2024lev} and use the framework to map to less-studied non-minimally coupled matter theories. While this framework was first developed primarily in the context of beyond-GR theories, it can in principle be extended to include any models of DM which modify the gravitational sector. Thus, we may be able to comment on initial constraints of possible DM candidates based on their interaction with gravity. However, we stress that our construction does not necessitate that the non-minimal couplings belong to a dark sector. Our parameterization is applicable to any matter sector, however, the possibility that DM experiences non-minimal couplings was primary motivation for this work. More generally, our main results are inspired by effective field theory (EFT) approaches to broad classes of non-minimally coupled theories.

The parameterizations in \cite{Jenks:2023pmk,Daniel:2024lev} have been shown to be consistent with and complementary to the parameterized post-Einsteinian (ppE) framework \cite{Yunes:2009ke,Cornish:2011ys,Tahura:2018zuq}, another novel tool with which to capture modifications of and deviations from GR. More recently, deep-learning neural network techniques have been utilized within the ppE framework \cite{Xie:2024ubm} to streamline computational efficiency and systematically explore the theory space in a sophisticated fashion (see also \cite{Wang:2024oei} for similar deep learning approaches in the post-Newtonian (PN) framework).

A particular modification in the gravitational sector would be the presence of gravitational birefringence, which manifests in the gravitational waveform if there are any parity-violating operators. Models including operators of this type have gained significant interest in recent years, the most studied being dynamical Chern-Simons gravity (dCS) \cite{Alexander:2009tp}. The sensitivities of third-generation GW detectors, such as Cosmic Explorer \cite{Evans:2021gyd} and the Einstein Telescope \cite{Maggiore:2019uih}, are expected to be good enough to be capable of detecting beyond-GR effects of this type \cite{Yunes:2024lzm}. Signatures of parity violation would also be present in the stochastic GW background as a net circular polarization \cite{Cruz:2024esk}. 

The upcoming space-based Laser Interferometer Space Antenna (LISA) \cite{LISA:2022yao} and the Chinese program Taiji \cite{Hu:2017mde} are also expected to be sensitive enough to detect parity-violating effects \cite{Chen:2024fto}, which have been predicted to have a wide variety of implications, from primordial GWs \cite{zhai2025primordialgravitationalwavesparityviolating} and a (chiral) GW background \cite{Orlando:2020oko,Chen:2024ikn,Su:2025nkl} to large-scale structure and inflation in cosmology \cite{Hewson:2024rnb, Hou:2024udn, Moretti:2024fzb, Reinhard:2024evr, Zhu:2024wme, Bao:2025onc}. It is also possible to detect parity violation within the stochastic background using pulsar polarization arrays  \cite{Liang:2025vji} and astrometry \cite{Liang:2023pbj,Jaraba:2025hay}. More generally, one needs to consider parity-even and parity-odd operators that lead to a combination of gravitational birefringence, modified dispersion relations, and perhaps Lorentz violation \cite{Qiao:2021fwi, Wang:2024erh, Allahyari:2025sbt, Khodadi:2025wuw, Lessa:2025kln, Wang:2025fhw, Zhang:2025kcw}.

The aforementioned framework provides an avenue for generally constraining extensions of GR and modified gravity theories. As previously mentioned, the GWs which have already been detected from LVK \cite{LIGOScientific:2014pky,VIRGO:2014yos}, in addition to pulsar timing array (PTA) experiments \cite{NANOGrav:2023gor,NANOGrav:2023hvm}, have thus far been in excellent agreement with GR, implying that conclusively observing beyond-GR effects will come down to precision physics. Furthermore, many beyond-GR theories have implications on astrophysical and cosmological systems when it comes to DM and beyond, from compact binary systems \cite{Aurrekoetxea:2024cqd, Fakhry:2024kjj, Mahmoudi:2024rga, Shen:2024qbb} and individual black holes or neutron stars \cite{Del_Popolo_2020, Duan:2023gng, Afrin:2024khy, Benetti:2024efb, Kehagias:2024yyp, Nashed:2024ppf, Baruah:2025ifh, Liu:2025fxj, Yang:2025byw, Pantig:2025eda, Yu:2025brr, Q:2025ycf}, to GW polarization and gravitational lensing \cite{Gao:2024ejs, Huang:2024gtu, Nomura:2024cku, Zhang:2025kze}, the cosmic microwave background (CMB) \cite{Mahmoudi:2024wjy, Smith:2024ayu}, inflation \cite{Fu:2024ipa, Hassan:2024vjg}, and broadly speaking, the dark sector \cite{Zhang2023, Bhattacharya:2024cpm, Bunji:2024ovg, Califano:2024xzt, Ghodla:2024fit, DeFelice:2025ykh, Figliolia:2025dtw, Zhang:2024bjo, Chen2025}. Indeed, theory-agnostic searches have gained popularity in recent years \cite{Crescimbeni:2024sam}. 

We stress that, in general, the task of probing beyond-GR effects in GWs is extremely challenging. It has been shown that in pure gravity (no non-minimal couplings to the matter sector), detectability of beyond-GR terms is not only highly restricted by solar system tests, but also by fundamental principles such as causality and unitarity \cite{Cassem:2024djm}. For example, the authors of \cite{Cassem:2024djm} found that quartic curvature terms in the gravitational sector can be probed if the current LIGO/Virgo precision increased by seven orders of magnitude, which perhaps could be prioritized in future ground-based interferometers. 

Additionally, utilizing the neural network/deep learning approach to ppE, recent work has shown that the GWTC-3 catalog is exceedingly consistent with GR \cite{Xie:2025voe}. Another ppE analysis was carried out in \cite{Zhao:2025mwq} suggesting strong evidence that black holes involved in producing LVK data are described by the Kerr geometry, further supporting the validity of GR. It is even possible for environmental effects or systematics to suggest a false violation of GR \cite{Garg:2024qxq} in GWs from mergers. 

Nonetheless, as outlined above, deviations from GR can be expressed in different systems\footnote{This includes, for example, the surprising phenomenon of demagnification in microlensing, as was recently explored in \cite{Zhang:2025kze} for the case of non-minimal DM couplings. Beyond GR effects on gravitational lensing have also been studied in \cite{Ezquiaga:2020dao} and more recently in \cite{Goyal:2023uvm}.}, and studying these extensions in various cosmological epochs can help to validate or rule out such beyond-GR theories. Additionally, exploring non-minimal DM couplings and DM distributions on various scales can shed new insight on fundamental physics in the early universe aside from gravity and DM. Therefore, it is important to be able to use GWs to probe beyond-GR theories throughout all of cosmic history, and this work aims to push forward our ability to do so.

With these motivations, the outline of this paper is as follows. After reviewing the existing parity-even and parity-odd parameterization in Section~\ref{sec:ParamEO}, we extend the parameterization in Section~\ref{sec:genericlagrangian} to account for early-universe effects and discuss its generic features. In Section~\ref{sec:ExampleModels} we apply the extended parameterization to three example models: Kalb-Ramond DM, axion-dilaton-Chern-Simons-Gauss-Bonnet, and a $U(1)$ (dark) photon model. We place constraints on the theory parameters for these models in Section~\ref{sec:constraints} using our parameterization that we have developed, before concluding in Section~\ref{sec:conclusion}. 

Throughout, we use units such that $m_p^2=\tfrac{1}{8\pi G} = c = 1$, although we occasionally restore the Planck mass explicitly. We assume a $(-,+,+,+)$ metric signature; Greek letters ($\mu$,$\nu$,...) range over all spacetime coordinates, Latin letters $(i,j,...)$ range over spatial indices, and square brackets denote anti-symmetrization over indices.

\section{Parity even and odd parameterization}\label{sec:ParamEO}

We define the gravitational perturbation $h_{\mu\nu}$ as is standard by
\begin{equation}
    g_{\mu\nu}=\bar{g}_{\mu\nu}+h_{\mu\nu}, \label{eq:graviton}
\end{equation}
where $\bar{g}_{\mu\nu}$ in this work will be the Friedmann-Lema\^itre-Robertson-Walker (FLRW) metric in conformal coordinates,
\begin{equation}
    d\bar{s}^2=a(\eta)^2\big(-d\eta^2+dx^2+dy^2+dz^2\big). \label{eq:flrw}
\end{equation}
Parity-violating effects in GW propagation are most easily seen by casting the linearized field equations in the left/right polarization basis. In the transverse-traceless (TT) gauge $\partial^\mu h_{\mu\nu}=h^\mu_\mu=0$, the gravitational perturbation is given explicitly by
\begin{equation}
    h_{\mu\nu}=\frac{1}{\sqrt{2}}\begin{pmatrix}
        0&0&0&0\\
        0&-(h_{\text{L}}+h_{\text{R}})&i(h_{\text{L}}-h_{\text{R}})&0 \\
        0&i(h_{\text{L}}-h_{\text{R}})&h_{\text{L}}+h_{\text{R}}&0 \\
        0&0&0&0 \label{eq:hmunu}
    \end{pmatrix},
\end{equation}
where the right- and left-handed strains are related to the usual $+,\times$ amplitudes by $h_{\text{R,L}}=\frac{1}{\sqrt{2}}(h_+\pm ih_\times).$ This parameterization assumes that the GW is propagating in the $z$-direction with polarizations in the $xy$-plane. 

Varying the Einstein-Hilbert Lagrangian with respect to each amplitude results in simple wave equations for each. Parity violation will then follow provided an additional interaction term sources the left- and right-handed amplitudes with a different sign, while parity-conserving modifications to GR appear with the same sign;
\begin{equation}\label{simpleEOM}
    \begin{split}
        \Box h_{\text{L}}&=+\Xi_O+\Xi_E,\\
        \Box h_{\text{R}}&=-\Xi_O+\Xi_E,
    \end{split}
\end{equation}
for some $\Xi_{O},\Xi_E$ that are expressions involving the GW strains and their derivatives, and the additional matter fields. 

\noindent
In general, the equations of motion take the form
\begin{equation}
    (1+A)h_{\text{R,L}}''+(2\mathcal{H}+B)h_{\text{R,L}}'+k^2(1+C)h_{\text{R,L}}=0,
\end{equation}
where $A,B,$ and $C$ stem from the beyond-GR couplings and satisfy $A,C\ll 1$, $|B|\ll \mathcal{H}$. We then divide $1+A$ from each term and expand to lowest order, $(1+A)^{-1}\approx 1-A,$ obtaining an equation of the form
\begin{equation} 
    h_{\text{R,L}}''+(2\mathcal{H}+\tilde{B})h_{\text{R,L}}'+k^2(1+\tilde{C})h_{\text{R,L}}=0.
    \end{equation}
Parity-even terms in $\tilde{B}$ or $\tilde{C}$ appear with the same sign for the right/left-handed strains, while the parity-odd terms appear with opposite signs. Any parity-violating terms in the coefficient $\tilde{B}$ produce \textit{amplitude birefringence}: one amplitude is attenuated while the opposite helicity amplitude is enhanced. A parity-violating term in $\tilde{C}$ produces \textit{velocity birefringence}\footnote{This effect is also  called \textit{phase birefringence}, since it appears in the GW solutions as a phase with opposing signs between the two helicities.}: the helicity amplitudes propagate with different dispersion relations and velocities. Any parity-even terms modify the amplitude and phase in the same fashion for each helicity. 

\noindent
The parameterization that was originally given \cite{Jenks:2023pmk}, and then generalized in  \cite{Daniel:2024lev}, is
\begin{align}\label{parameterization}
    &h_{R,L}''~+~\bigg\{2\mathcal{H}~+~\sum_{n=0}^{\infty}(\lambda_{\text{R,L}}k)^n\bigg[\frac{\alpha_n(\eta)}{(\Lambda a)^n}\mathcal{H}~+~\frac{\beta_n(\eta)}{(\Lambda a)^{n-1}}\bigg]\bigg\}h_{\text{R,L}}' \nonumber \\ +~&k^2\bigg\{1~+~\sum_{m=0}^{\infty}(\lambda_{\text{R,L}})^{m+1}k^{m-1}\bigg[\frac{\gamma_m(\eta)}{(\Lambda a)^m}\mathcal{H}~+~\frac{\delta_m(\eta)}{(\Lambda a)^{m-1}}\bigg]\bigg\}h_{\text{R,L}} = 0, 
\end{align}
where primes denote partial derivatives with respect to conformal time $\eta$, $\mathcal{H}\equiv a'/a$ is the Hubble parameter, $\lambda_{\text{R,L}}=\pm 1,$ and $\Lambda$ is the EFT cutoff. The summation indices here run over all integers, and the $\{\alpha_n,\beta_n,\gamma_m,\delta_m\}$ functions generically depend on the conformal time and are theory-dependent. In what follows, we will write down a more general parameterization which captures effects at $O(\mathcal{H}^2)$ and $O(\mathcal{H}')$.

\section{General Lagrangian and the extended parameterization}\label{sec:genericlagrangian}
Our focus in this work will be on parity-even and parity-odd non-minimal couplings between curvature and the matter sector, with fields collectively denoted as $\Psi$. In general, such models can be written as
\begin{equation}
    S=\int d^4x\sqrt{-g}\Big(\tfrac{1}{2}R+\mathcal{L}_\Psi-\mathcal{L}_{int}+\mathcal{L}_m\Big),
\end{equation}
where $\mathcal{L}_m$ is the minimally coupled matter content, and the interaction Lagrangian is 
\begin{equation}\label{genericLagrangian}
    \begin{split} \mathcal{L}_{int}&=\tilde{R}^{\mu\nu\rho\sigma}\sum_{m=1}\frac{\tilde{\xi}_m}{(\Lambda_O)^{\tilde{\zeta}_m}}\mathcal{O}_{\mu\nu\rho\sigma}^m(\Psi,g)+R^{\mu\nu\rho\sigma}\sum_{n=1}\frac{\xi_n}{(\Lambda_E)^{\zeta_n}}\mathcal{E}_{\mu\nu\rho\sigma}^n(\Psi,g).
    \end{split}
\end{equation}
In (\ref{genericLagrangian}), $\xi$ and $\tilde{\xi}$ are dimensionless constants, the $\Lambda_E,\Lambda_O$ are the effective cutoffs for the parity-even and parity-odd interactions, respectively, and the $\zeta, \ \tilde{\zeta}$ are integers that account for the mass dimension. The tensors $\mathcal{E}_{\mu\nu\rho\sigma}$ and $\mathcal{O}_{\mu\nu\rho\sigma}$ are in general functions of the matter fields and their derivatives, as well as possibly the Riemann curvature and its contractions. The presence of the dual Riemann tensor $\tilde{R}^{\mu\nu\rho\sigma}=\frac{1}{2\sqrt{-g}}\varepsilon^{\mu\nu\alpha\beta}R_{\alpha\beta}^{ \ \ \ \rho\sigma}$ generally results in the equations of motion for the left-handed strains having the opposite sign of the same term to the right-handed strains, analogous to the $\Xi_O$ term in (\ref{simpleEOM}).

\noindent
We recast the parameterization (\ref{parameterization}) in the following way:
 \begin{equation}\label{GenParam}
 \begin{split} 
&  h''_{\text{R,L}}+\Big\{2\mathcal{H}+C^{(1)}_O+C^{(1)}_E\Big\}h'_{\text{R,L}} +k^2\Big\{1+C^{(0)}_O+C^{(0)}_E\Big\} h_{\text{R,L}}=0, 
 \end{split}
 \end{equation}
where we have split the parity-even and -odd contributions in the first and zeroth derivative terms. We will moreover allow different EFT cutoffs for parity-even versus -odd operators, $\Lambda_E$ and $\Lambda_O$, as in (\ref{genericLagrangian})\footnote{From a phenomenological perspective, allowing for different EFT cutoffs enables one to capture effects from opposite parity operators that may appear at different scales.}. We define 
    \begin{equation}\label{COsCEs}
        \begin{split}
            C^{(1)}_O&=\lambda_{\text{R,L}}\sum_{n=0}k^{2n+1}\Bigg[\frac{\alpha_{2n+1}}{(\Lambda_O a)^{2n+1}}\mathcal{H}+\frac{\beta_{2n+1}}{(\Lambda_O a)^{2n}}+\frac{\mu_{2n+1}\mathcal{H}^2 + \rho_{2n+1}\mathcal{H}'}{(\Lambda_Oa)^{2n+2}}\Bigg], \\
        C^{(1)}_E&=\sum_{n=0}k^{2n}\Bigg[\frac{\alpha_{2n}}{(\Lambda_E a)^{2n}}\mathcal{H}+\frac{\beta_{2n}}{(\Lambda_E a)^{2n-1}}+\frac{\mu_{2n}\mathcal{H}^2 + \rho_{2n}\mathcal{H}'}{(\Lambda_Ea)^{2n+1}}\Bigg], \\
        C^{(0)}_O& 
        =\lambda_{\text{R,L}}\sum_{m=0}k^{2m-1}\Bigg[\frac{\gamma_{2m}}{(\Lambda_O a)^{2m}}\mathcal{H}+\frac{1}{(\Lambda_O a)^{2m-1}}\Big(\delta_{2m}+\frac{\nu_{2m}\mathcal{H}^2 + \sigma_{2m}\mathcal{H}'}{k^2}\Big)\Bigg], \\
        C^{(0)}_E&=\sum_{m=0}k^{2m}\Bigg[\frac{\gamma_{2m+1}}{(\Lambda_E a)^{2m+1}}\mathcal{H}+\frac{1}{(\Lambda_E a)^{2m}}\Big(\delta_{2m+1}+\frac{\nu_{2m+1}\mathcal{H}^2 + \sigma_{2m+1}\mathcal{H}'}{k^2}\Big)\Bigg], 
        \end{split}
    \end{equation}
where the functions $\{\alpha_n,\beta_n,\gamma_n,\delta_n,\mu_n,\nu_n,\rho_n,\sigma_n\}$ are all theory-dependent and $\beta_0=\delta_0=\nu_0=\sigma_0=0$. 

Eq.~(\ref{GenParam}) provides a systematic EFT-like parameterization of the parity-odd sector, incorporating both amplitude and phase birefringence. This differs from earlier studies, which either treated parity-violating effects schematically or focused on specific theories such as Chern-Simons gravity \cite{Alexander:2009tp,Nishizawa:2018srh}. More general parameterizations of GW propagation have been developed in the parity-even sector, and within the 
effective field theory of dark energy (EFT-of-DE), where the coefficients are typically assumed to only depend on time \cite{Gubitosi2013,Bellini:2014fua,Bloomfield2013,Ezquiaga2017,LISACosmologyWorkingGroup:2019mwx}\footnote{Specifically, the canonical cosmological modified gravity tensor equation is 
\begin{equation}
    {h}''_{ij} + (2 + \alpha_M)\mathcal{H}h'_{ij} + c_T^2k^2h_{ij} = 0, \label{eq:mod-grav-tensor-eq}
\end{equation}
which is the Bellini-Sawicki / EFT-of-DE parameterization \cite{Bellini:2014fua,Amendola:2017ovw}. After field redefinitions, this is equivalent to earlier propagation frameworks \cite{Nishizawa:2017nef,Ezquiaga2021} which capture the parity-even sector as 
\begin{equation}
    h'' + (2 + \nu)\mathcal{H}h' + c_T^2k^2h + \mu^2h = 0, 
\end{equation}
with $\nu \leftrightarrow \alpha_M$ and $c_T^2 \leftrightarrow 1 + \alpha_T$.}. In contrast, we allow the coefficients to depend on both time and momentum, thereby capturing higher-derivative operators, Lorentz-violating dispersion, and other UV-sensitive effects \cite{Mirshekari_2012,Kostelecky2016}. By treating modified friction, dispersion, parity violation and momentum dependence within a single propagation equation, Eq.~(\ref{GenParam}) can be directly mapped onto waveform-level parameterizations such as the ppE framework \cite{Yunes:2009hc,Cornish:2011ys}.

It is straightforward to see by inspection that the $\mathcal{H}^2$ and $\mathcal{H}'$ terms have the correct form. We  require the correct dimensionality as well as consistency with the flat space limit, $\mathcal{H}\rightarrow 0$, and the low-energy limit $k\rightarrow 0$, namely, that all beyond-GR terms are finite.

\noindent
Combining the momenta dependence in each sum, we see that 
\begin{equation}
    \begin{split}
        k^{2n+1}\frac{\mu_{2n+1}}{(\Lambda_Oa)^{2n+2}}\mathcal{H}^2, \ \ \ \ \ \ \ 
        k^{2n}\frac{\mu_{2n}}{(\Lambda_Ea)^{2n+1}}\mathcal{H}^2
    \end{split}
\end{equation}
both have mass dimension one for all $n$, as required, and
\begin{equation}
    \begin{split}
      k^2  k^{2m-1}\frac{\nu_{2m}}{(\Lambda_Oa)^{2m-1}}\Big(\frac{\mathcal{H}}{k}\Big)^2, \ \ \ \ \ \ \ \ 
      k^2k^{2m}\frac{\nu_{2m+1}}{(\Lambda_Ea)^{2m}}\Big(\frac{\mathcal{H}}{k}\Big)^2
    \end{split}
\end{equation}
both have mass dimension two for all $m$, as required. The parity-violating terms straightforwardly expand with odd powers of $k$, while the parity-conserving terms expand with even powers of $k$.

We here note a few assumptions that are made in order for us to focus on a propagation equation of the form of \eqref{GenParam}, leading to the solution \eqref{eq:hrlgenmod}. In general, theories of the form of \eqref{genericLagrangian} can potentially propagate higher time derivatives, which are not accounted for in \eqref{GenParam}. For a given operator, an $n^{th}$ order time derivative contributes to the equations of motion a term of the form $\sim\omega^n/\Lambda^m,$ where $m$ accounts for for the mass dimension. We restrict ourselves to the parametric regime where $\omega^n\ll\Lambda^m,$ so that such terms are subdominant.

It is also generally the case that non-minimally coupled theories propagate additional degrees of freedom in the metric perturbations. Our work here is to focus on the effect of non-minimal couplings on the physical, TT sector in order to construct a template that is ultimately used to compare to interferometer data. Indeed, mixing between the TT GW and additional perturbative degrees of freedom may lead to interesting observational signatures. We assume here that such dynamics are negligible and focus solely on the background non-minimally coupled matter sector's affect on the GWs, and leave this interesting subtlety for future work.

We additionally assume that any fluctuations in the matter sector are further subdominant, \textit{i.e.} $\delta\alpha\ll\alpha_0,$ for all fields under consideration. Relatedly, we focus on an isotropic background. As we will discuss in section \ref{sec:KRDM}, it is possible for the matter sector to attain directional expectation values which may break isotropy, at which point the appropriate cosmological background will be of the Bianchi type. Our parameterization necessitates that any such terms are completely negligible when compared to the dominating matter sector that drives isotropic expansion. It would be interesting to generalize \eqref{GenParam} to anisotropic cosmologies, which is beyond the scope of this work.

From the propagation equations, we can find the explicit corrections to $h_{\text{R,L}}$. The right- and left-handed modes are modified in the following way\footnote{See Appendix \ref{sec:AppendixA:GenProp} for full derivation.}:
\begin{align}
    h_{\text{R,L}} &= \Bar{h}_{\text{R,L}} \nonumber \\ &\times \text{exp}\bigg\{-(\lambda_{\text{R,L}})^{2n}\bigg[\frac{[k(1+z)]^{2n}}{2}\bigg(\frac{\alpha_{2n_0}}{\Lambda_E^{2n}}z_{2n} + \frac{\beta_{2n_0}}{\Lambda_E^{2n-1}}D_{2n+1} + \frac{\mu_{2n_0}}{\Lambda_E^{2n+1}}\int Hdz \nonumber \\ &+ \frac{\rho_{2n_0}}{\Lambda_E^{2n+1}}\int (1+z)^{-1}H_zdz\bigg) + \lambda_{R,L}\frac{[k(1+z)]^{2n+1}}{2}\bigg(\frac{\alpha_{2n_0+1}}{\Lambda_O^{2n+1}}z_{2n+1} + \frac{\beta_{2n_0+1}}{\Lambda_O^{2n}}D_{2n+2} \nonumber \\ &+ \frac{\mu_{2n_0+1}}{\Lambda_O^{2n+2}}\int Hdz + \frac{\rho_{2n_0+1}}{\Lambda_O^{2n+2}}\int(1+z)^{-1}H_zdz\bigg)\bigg]\bigg\} \nonumber \\ &\times \text{exp}\bigg\{i(\lambda_{\text{R,L}})^{2m+1}\bigg[\frac{[k(1+z)]^{2m}}{2}\bigg(\frac{\gamma_{2m_0}}{\Lambda_O^{2m}}z_{2m} + \frac{\delta_{2m_0}}{\Lambda_O^{2m-1}}D_{2m+1} + \frac{\nu_{2m_0}}{\Lambda_O^{2m-1}k^2}\int Hdz \nonumber \\ &+ \frac{\sigma_{2m_0}}{\Lambda_O^{2m-1}k^2}\int (1+z)^{-1}H_zdz\bigg) + \lambda_{\text{R,L}}\frac{[k(1+z)]^{2m+1}}{2}\bigg(\frac{\gamma_{2m_0+1}}{\Lambda_E^{2m+1}}z_{2m+1} + \frac{\delta_{2m_0+1}}{\Lambda_E^{2m}}D_{2m+2} \nonumber \\ &+ \frac{\nu_{2m_0+1}}{\Lambda_E^{2m}k^2}\int Hdz + \frac{\sigma_{2m_0+1}}{\Lambda_E^{2m}k^2}\int (1+z)^{-1}H_zdz\bigg)\bigg]\bigg\}, \label{eq:hrlgenmod}
\end{align}
where $\Bar{h}_{\text{R,L}}$ is the GR solution for the right- and left-handed modes, the ``0" subscript indicates the present-day value, $\int_0^z H(z')dz'$ can be integrated numerically using that $H(z) = H_0\sqrt{\Omega_{m,0}(1+z)^3 + \Omega_{r,0}(1+z)^4 + \Omega_{\Lambda,0}}$, and $H_z \equiv dH(z)/dz$, with $H \equiv \mathcal{H}/a$ being the cosmic Hubble parameter. The summation over $m$ and $n$ is implied, and we define an effective distance, $D_{\alpha}$ \cite{Mirshekari_2012, Ezquiaga_2022},
\begin{align}
    D_{\alpha} = (1+z)^{1-\alpha}\int\frac{(1+z)^{\alpha - 2}}{H(z)}dz,
\end{align}
as well as an effective redshift parameter, $z_{\alpha}$, such that\footnote{We note that $D_1 = D_T$, where $D_T$ is the look-back distance, and $D_2 = (1+z)^{-1}D_C = D_A$, where $D_C$ and $D_A$ are the comoving and angular-diameter distances, respectively. We observe that $z_0 = \text{ln}(1+z)$ and $z_1 = z(1+z)^{-1}$.}
\begin{align}
    z_{\alpha} = (1+z)^{-\alpha}\int\frac{dz}{(1+z)^{1-\alpha}}.
\end{align}

Eq.~(\ref{eq:hrlgenmod}) assumes a WKB (geometric optics) regime in which the GW frequency is large compared to the background variation scale, $\omega \gg H$, ensuring adiabatic evolution of both amplitude and phase. At the same time, since the modified dispersion relations arise within an EFT, the frequency and physical wavenumber must remain below the cutoff scale ($\omega, k/a \ll \Lambda)$, as noted earlier. Therefore, the controlled regime of Eq.~(\ref{eq:hrlgenmod}) is characterized by the parametric window $H \ll \omega \ll \Lambda$.

\noindent
From Eqs.~(\ref{GenParam}) and (\ref{COsCEs}), we can read off the modified dispersion relation,
\begin{equation}\label{gendispersion}
\begin{split}
    \omega_{\text{R,L}}^2 = k^2\bigg\{1 &+ \lambda_{\text{R,L}}k^{2m-1}\bigg[\frac{\gamma_{2m}}{(\Lambda_Oa)^{2m}}\mathcal{H} + \frac{1}{(\Lambda_Oa)^{2m-1}}\bigg(\delta_{2m} + \frac{\nu_{2m}\mathcal{H}^2 + \sigma_{2m}\mathcal{H}'}{k^2}\bigg)\bigg]  \\ &+ k^{2m}\bigg[\frac{\gamma_{2m+1}}{(\Lambda_Ea)^{2m+1}}\mathcal{H} + \frac{1}{(\Lambda_Ea)^{2m}}\bigg(\delta_{2m+1} + \frac{\nu_{2m+1}\mathcal{H}^2 + \sigma_{2m+1}\mathcal{H}'}{k^2}\bigg)\bigg]\bigg\}.
    \end{split}
\end{equation}
Thus, to lowest order, the modified group and phase velocities are 
\begin{equation}\label{gengroupvelocity}
\begin{split} 
    v_g^{\text{R,L}} &= 1 + \lambda_{\text{R,L}}k^{2m-1}m\bigg[\frac{\gamma_{2m}}{(\Lambda_Oa)^{2m}}\mathcal{H} + \frac{1}{(\Lambda_Oa)^{2m-1}}\bigg(\delta_{2m} + \frac{\nu_{2m}\mathcal{H}^2 + \sigma_{2m}\mathcal{H}'}{k^2}\bigg)\bigg]  \\ &+ k^{2m}\bigg(m + \frac{1}{2}\bigg)\bigg[\frac{\gamma_{2m+1}}{(\Lambda_Ea)^{2m+1}}\mathcal{H} + \frac{1}{(\Lambda_Ea)^{2m}}\bigg(\delta_{2m+1} + \frac{\nu_{2m+1}\mathcal{H}^2 + \sigma_{2m+1}\mathcal{H}'}{k^2}\bigg)\bigg] \\ &- \lambda_{\text{R,L}}k^{2m-1}\frac{\nu_{2m}\mathcal{H}^2 + \sigma_{2m}\mathcal{H}'}{(\Lambda_Oa)^{2m-1}k^2}  -k^{2m}\bigg[\frac{\nu_{2m+1}\mathcal{H}^2 + \sigma_{2m+1}\mathcal{H}'}{(\Lambda_Ea)^{2m}k^2}\bigg],
\end{split}
\end{equation}
\begin{equation}\label{genphasevelocity}
\begin{split}
    v_p^{\text{R,L}} &= 1 + \frac{\lambda_{\text{R,L}}k^{2m-1}}{2}\bigg[\frac{\gamma_{2m}}{(\Lambda_Oa)^{2m}}\mathcal{H} + \frac{1}{(\Lambda_Oa)^{2m-1}}\bigg(\delta_{2m} + \frac{\nu_{2m}\mathcal{H}^2 + \sigma_{2m}\mathcal{H}'}{k^2}\bigg)\bigg] \\ 
    &+ \frac{k^{2m}}{2}\bigg[\frac{\gamma_{2m+1}}{(\Lambda_Ea)^{2m+1}}\mathcal{H} + \frac{1}{(\Lambda_Ea)^{2m}}\bigg(\delta_{2m+1} + \frac{\nu_{2m+1}\mathcal{H}^2 + \sigma_{2m+1}\mathcal{H}'}{k^2}\bigg)\bigg].
\end{split}
\end{equation}
Although the solution, dispersion relation, group and phase velocities appear cumbersome and overly complicated in this formalism, in practice, models of interest have very few total terms.

\subsection{Generic features}\label{sec:GenericFeatures}

The presence of the non-minimal couplings result in the GW equations of motion exhibiting some interesting features in the two limits: the superhorizon limit $k\ll\mathcal{H}$ and a Minkowski-like limit\footnote{In the strict Minkowski limit $\mathcal{H}\rightarrow0$, no $\beta$-terms may survive, as Poincar\'e invariance is restored such that no preferred timelike vector can is available to contract to produce a single time derivative. Our limit here should be thought of as the limit where all modes are deep inside the horizon.}, $k\gg\mathcal{H},k\beta\gg\mathcal{H}$. In the latter limit, the $\beta$ and $\delta$ functions dominate. In the simple example where $\beta_1,\beta_2,\delta_1,\delta_2$ are nonzero, the equations of motion reduce to 
\begin{equation}
    h_{R,L}''+\Big(2\mathcal{H} + \lambda_{R,L}k\beta_1+k^2\frac{\beta_2}{\Lambda_E}\Big)h_{R,L}'+k^2\Big(1+\delta_1+\lambda_{R,L}k\frac{\delta_2}{\Lambda_O}\Big)h_{R,L}=0.
\end{equation}
Assuming the fields are slowly varying, this is solved (in coordinate time) by
\begin{equation}
    h_{R,L}=h_{R,L}^{(0)}\text{exp}\Big[(X^k_{R,L}+i\omega^k_{R,L})t\Big],
\end{equation}
where to lowest order in the beyond-GR parameters,
\begin{equation}
    X^k_{R,L}=-\frac{1}{2}\Big(\lambda_{R,L}k\beta_1+k^2\frac{\beta_2}{\Lambda_E}\Big), \ \ \ \ \omega^k_{R,L}=k\Big(1+\frac{1}{2}\delta_1+\lambda_{R,L}k\frac{\delta_2}{2\Lambda_O}\Big).
\end{equation}
We see amplitude and velocity birefringence from the parity-odd terms, a modified dispersion relation, and momentum-dependent amplitude enhancement/attenuation. 

Let us now consider the $k\rightarrow 0$ limit\footnote{No physical GW will have $k=0$ exactly. More specifically, this limit corresponds to focusing on GWs where the wavelength satisfies $\lambda\gg H^{-1}$.} in a cosmological background, where only a few parity-even terms are relevant. The equation of motion for both helicity amplitudes is 
\begin{equation}\label{kto0EOM}
    h''+\Big(2\mathcal{H}+\alpha_0\mathcal{H}+\frac{\mu_0\mathcal{H}^2+\rho_0\mathcal{H}'}{\Lambda_Ea}\Big)h'+(\nu_1\mathcal{H}^2+\sigma_1\mathcal{H}')h=0.
\end{equation}
During inflation in de Sitter space, $a(\eta)=-(H_0\eta)^{-1}$ and $\mathcal{H}=-\eta^{-1}$, where $H_0$ is the Hubble constant, we have
\begin{equation}
    h''-\frac{1}{\eta}\Big[2+\alpha_0+\frac{H_0}{\Lambda_E}(\mu_0+\rho_0)\Big]h'+\frac{1}{\eta^2}\Big(\nu_1+\sigma_1\Big)h=0.
\end{equation}
If the beyond-GR parameters are slowly varying\footnote{This assumption possibly breaks down in realistic early-universe scenarios, which we discuss further in Sec. \ref{sec:conclusion}.}, this is solved by 
\begin{equation}
    h(\eta)=c_0\eta^{x_1}+c_1\eta^{x_2},
\end{equation}
for integration constants $c_0,c_1$, with 
\begin{equation}
    x_1=\nu_1+\sigma_1, \ \ \ \ x_2=3+\alpha_0-\nu_1-\sigma_1+\frac{H_0}{\Lambda_E}(\mu_0+\rho_0)
\end{equation}
at lowest order. The solutions are easier to interpret in coordinate time $t$, related to conformal time in de Sitter as $\eta\sim H_0^{-1}e^{-H_0t}$. Here, we have
\begin{equation}
    h(t)=\tilde{c}_0e^{-(H_0x_1)t}+\tilde{c}_1e^{-(H_0x_2)t}.
\end{equation}
If we turn off all beyond-GR terms, we obtain the standard superhorizon modes that are constant plus a decaying piece, $h(t)=\tilde{c}_0+\tilde{c}_1e^{-3H_0t}$. This will always be the dominant contribution to the solution, but more generally, the beyond-GR terms result in slight modifications to the ultra-long wavelength modes. 

In this long-wavelength limit $k\rightarrow0$, tensor perturbations correspond to large-scale metric deformations that are locally indistinguishable from coordinate transformations. As a consequence of diffeomorphism invariance, the equations of motion must remain regular in this limit and cannot generate independent propagating dynamics for these modes. In our parameterization, this is reflected by the absence of singular terms in $k$ and by the fact that the leading behavior is controlled entirely by background quantities such as $\mathcal{H}$ and $\mathcal{H}'$, as is clear from \eqref{kto0EOM}, reproducing the expected behavior.

\section{Example models}\label{sec:ExampleModels}

\subsection{Kalb-Ramond dark matter}\label{sec:KRDM}
It was first pointed out in \cite{Capanelli:2023uwv} that Kalb-Ramond-like particles could be a viable DM candidate. Moreover, two of the current authors studied in \cite{Manton:2024hyc} a dimension-four operator coupling the dual Riemann tensor to the Kalb-Ramond (KR) 2-form field $B^{\mu\nu}$. We will additionally include all allowed dimension-four operators, and consider the Lagrangian
\begin{equation}\label{LintKR}
\begin{split} 
\mathcal{L}_{int}&=\tilde{\xi}_1\tilde{R}^{\mu\nu\rho\sigma}B_{\mu\nu}B_{\rho\sigma}+\xi_1 R^{\mu\nu\rho\sigma}B_{\mu\nu}B_{\rho\sigma}+\xi_2R^{\mu\rho}B_{\mu\lambda}B_{ \ \rho}^{ \lambda}+\xi_3RB_{\mu\nu}B^{\mu\nu}+\xi_4RB_{\mu\nu}\tilde{B}^{\mu\nu},
\end{split} 
\end{equation}
where $\tilde{B}^{\mu\nu}=\frac{1}{2}\epsilon^{\mu\nu\rho\sigma}B_{\rho\sigma}$ is the dual. 

The couplings $\xi_3,\xi_4$ are of the type $\bar{f}(R)\mathcal{L}_m$, which was studied in the context of $f(R)$ gravity in \cite{Bertolami:2017svl}. Furthermore, the model from \cite{Manton:2024hyc} is recovered in the limit $\tilde{\xi}_1\rightarrow\tilde{\xi}_1/2,$ $\xi_1,\xi_2,\xi_3,\xi_4\rightarrow 0$. Note that there is an additional dimension-four operator $\mathcal{L}\sim\xi_5 R^{\mu\nu}B_{\mu\lambda}\tilde{B}^{\lambda}_{ \ \nu}$; however, the contribution from this term can be absorbed into the constant $\xi_4$ using the identity $B_{\mu\lambda}\tilde{B}^\lambda_{ \ \nu}=\frac{1}{4}g_{\mu\nu}B_{\alpha\beta}\tilde{B}^{\alpha\beta}$.

In the notation of (\ref{genericLagrangian}), since the operators are dimension-four, we have $\zeta=\tilde{\zeta}=0$ for all terms, 
\begin{equation}
    \mathcal{O}^1_{\mu\nu\rho\sigma}=\mathcal{E}^1_{\mu\nu\rho\sigma}=B_{\mu\nu}B_{\rho\sigma},
\end{equation}
for the contractions with the Riemann tensor and its dual, along with 
\begin{equation}
    \mathcal{E}^2_{\mu\nu\rho\sigma}=g_{\nu\sigma}B_{\mu\lambda}B^\lambda_{ \ \nu}, \ \ \ \ \ \ \mathcal{E}^3_{\mu\nu\rho\sigma}=g_{\mu\rho}g_{\nu\sigma}B_{\alpha\beta}B^{\alpha\beta}, \ \ \ \ \ \ \ \mathcal{E}^4_{\mu\nu\rho\sigma}=g_{\mu\rho}g_{\nu\sigma}B_{\alpha\beta}\tilde{B}^{\alpha\beta}.
\end{equation}
The KR field is described by the Lagrangian \cite{PhysRevD.9.2273} 
\begin{equation}\label{LB}
    \mathcal{L}_{\text{B}}=\frac{1}{12}H_{\mu\nu\rho}H^{\mu\nu\rho}-V(B),
\end{equation}
where 
\begin{equation} 
H_{\mu\nu\rho}=\partial_{[\mu}B_{\nu\rho]}=\partial_\mu B_{\nu\rho}+\partial_\nu B_{\rho\mu}+\partial_\rho B_{\mu\nu}
\end{equation}
is the KR field strength and $V(B)$ is a potential. The non-interacting theory with $V=0$ admits the following gauge symmetries\footnote{We follow the anti-symmetrization convention $A_{[\mu}B_{\nu]}=A_\mu B_\nu-A_\nu B_\mu$ and symmetrization $A_{(\mu}B_{\nu)}=A_\mu B_\nu+A_\nu B_\mu$.}  
\begin{equation}\label{gauge1}
    B_{\mu\nu}\rightarrow B_{\mu\nu}+\partial_{[\mu}b_{\nu]} 
\end{equation}
for an arbitrary vector $b_\mu(x)$, and
\begin{equation}\label{gauge2}
    b_\mu\rightarrow b_\mu+\partial_\mu\lambda,
\end{equation}
for an arbitrary scalar function $\lambda(x)$. As is well known, the KR field strength in four dimensions is dual to a pseudoscalar $\alpha$ as $\partial_\mu\alpha\sim\varepsilon_{\mu\nu\rho\sigma}H^{\nu\rho\sigma}$, which is often referred to as the KR axion. However in the case of broken gauge invariance by $V(B)\neq 0,$ the field strength can not be dualized and $B_{\mu\nu}$ is the propagating degree of freedom.

\noindent
The full theory is $\mathcal{L}=\mathcal{L}_{EH}+\mathcal{L}_B-\mathcal{L}_{int}+\mathcal{L}_m$, with equations of motion
\begin{equation}
    \frac{1}{2}\nabla_\alpha H^\alpha_{ \ \ \mu\nu}=-\frac{\partial V}{\partial B^{\mu\nu}}-2\xi_1R_{\mu\nu\rho\sigma}B^{\rho\sigma}-\xi_2R_{\alpha[\mu}B_{\nu]}^{ \ \alpha}-2\xi_3RB_{\mu\nu}-2\xi_4R\tilde{B}_{\mu\nu}-2\tilde{\xi}_1\tilde{R}_{\mu\nu\rho\sigma}B^{\rho\sigma}
\end{equation}
for the KR field, and
\begin{equation}\label{KRgravEquations}
    \begin{split}
        m_p^2G_{\mu\nu}&=\tilde{\Theta}_{\mu\nu}+\sum_{i=1}^4\Theta^{(i)}_{\mu\nu}+\frac{1}{2}g_{\mu\nu}\mathcal{L}_{int}+T_{\mu\nu}^{\text{B}}+T^m_{\mu\nu},
    \end{split}
\end{equation}
where $T^m_{\mu\nu}$ is the stress-tensor for the minimally coupled matter content and
\begin{equation}\label{KRTmunu}
     T_{\mu\nu}^{\text{B}}=-\frac{1}{2}H_\mu^{ \ \ \alpha\beta}H_{\nu\alpha\beta}+g_{\mu\nu}\Big(\frac{1}{12}H_{\alpha\beta\rho}H^{\alpha\beta\rho}-V(B)\Big)
\end{equation}
is the contribution of the non-interacting part of the KR field. Now, the first term from the non-minimal couplings stems from the contraction with the dual Riemann tensor,
\begin{equation}
    \tilde{\Theta}_{\mu\nu}=2\tilde{\xi}_1\nabla^\alpha\nabla^\beta\tilde{P}_{\mu\alpha\nu\beta},
\end{equation}
and we have defined the shorthand 
\begin{equation}\label{Ptilde}
    \tilde{P}^{\mu\nu\rho\sigma}=\frac{1}{3}\Big(\tBk^{\mu\nu}B^{\rho\sigma}+B^{\mu\nu}\tBk^{\rho\sigma}\Big) +\frac{1}{6}\Big(\tBk^{\mu\rho}B^{\nu\sigma}+B^{\mu\rho}\tBk^{\nu\sigma}+\tBk^{\rho\nu}B^{\mu\sigma}+B^{\rho\nu}\tBk^{\mu\sigma}\Big).
\end{equation}
Note that this combination possesses all the symmetries of the Riemann tensor. For the parity-even pieces, we similarly have
\begin{equation}
    \Theta^{(1)}_{\mu\nu}=2\xi_1\nabla^\alpha\nabla^\beta P_{\mu\alpha\nu\beta},
\end{equation}
where $P^{\mu\nu\rho\sigma}$ is analogous to (\ref{Ptilde}) except no dualized KR fields. Next, we have
\begin{equation}
    \Theta^{(2)}_{\mu\nu}=\frac{\xi_2}{2}\Big(\nabla_\alpha\nabla_{(\mu}[B_{\nu)}^{ \ \lambda}B_\lambda^{ \ \alpha}]-\Box B_\mu^{ \ \lambda}B_{\lambda\nu}-g_{\mu\nu}\nabla_\alpha\nabla_\beta(B^{\alpha\lambda}B_\lambda^{ \ \beta})-R_{\alpha(\mu}B_{\nu)\lambda}B^{\lambda\alpha}\Big)
\end{equation}
stemming from the coupling to the Ricci tensor, and finally
\begin{equation}
    \Theta^{(3)}_{\mu\nu}+\Theta^{(4)}_{\mu\nu}=\Big(\nabla_\mu\nabla_\nu-g_{\mu\nu}\Box-R_{\mu\nu}\Big)\Big(\xi_3B_{\alpha\beta}B^{\alpha\beta}+\xi_4B_{\alpha\beta}\tilde{B}^{\alpha\beta}\Big)
\end{equation}
stems from the two couplings to the Ricci scalar.

As was studied in \cite{Manton:2024hyc}, we will here consider a simplified scenario where the components of $B_{\mu\nu}$ take on constant background values $B^2=B_{(0)}^2$, which we model by a symmetry breaking potential \begin{equation}\label{potential}
    V(B_{\mu\nu}B^{\mu\nu}-v^2).
\end{equation}
The KR field settles into the minimum of the potential where
\begin{equation}
    v^2\rightarrow \bar{g}^{\mu\alpha}\bar{g}^{\nu\beta}B_{\mu\nu}^{(0)}B_{\alpha\beta}^{(0)}.
\end{equation}
Effective potentials that possess a stable minima can arise from quantum corrections due to fermion couplings \cite{Assuncao:2019azw}. Potentials of this form in general break Lorentz invariance, as is discussed at length in \cite{Altschul:2009ae,Malta:2025ydq} and elsewhere, although this is not important for our current purposes. 

\noindent
The covariant tensor can be parameterized\footnote{When cosmological evolution is dominated by two-form fields, one either needs to introduce multiple fields \cite{Kobayashi:2009hj,Germani:2009iq} analogously to vector inflation \cite{Golovnev:2008cf} to preserve isotropy, or the cosmological evolution behaves as an anisotropic Bianchi universe \cite{Maluf:2021eyu}. We approximate our background as isotropic, therefore assuming the KR matter sector does not significantly contribute to cosmological expansion. } as \cite{Manton:2024hyc,Altschul:2009ae,Maluf:2021eyu}
\begin{equation}\label{vev}
    B^{(0)}_{\mu\nu}=\frac{1}{\sqrt{2}}\begin{pmatrix}
        0&-\Ve a^2&0&0\\
        \Ve a^2 &0&0&0 \\
        0&0&0&\Vm a^2 \\
        0&0&-\Vm a^2 & 0
    \end{pmatrix},
\end{equation}
where $\Ve$ and $\Vm$ are real numbers of mass dimension one, and $a$ is the conformal scale factor. We then have that the minimum of the potential is $v=\sqrt{\Vm^2-\Ve^2}$. 

\noindent
The equation of motion for the GW amplitudes simplifies to
\begin{equation}\label{hEOMKR1}
    h_{R,L}''+2\mathcal{H}h_{R,L}'+k^2\Big(1+A+\lambda_{R,L}B\frac{\mathcal{H}}{k}+C\frac{(\mathcal{H}'+2\mathcal{H}^2)}{k^2}\Big)h_{R,L}=0,
\end{equation}
where
\begin{equation}
    \begin{split}
        A&=(\tVm^2+\tVe^2)(\xi_2-6\xi_1), \\
        B&=2\Big(2\tVe\tVm\xi_1-\tilde{\xi}_1(\tVm^2-\tVe^2)\Big), \\
        C&=2\Big(\tVe^2(2\xi_1-2\xi_3)+\tVm^2(2\xi_1+2\xi_3-\xi_2)+4\xi_4\tVe\tVm\Big),
    \end{split}
\end{equation}
restoring the Planck mass and defining the dimensionless quantities $\tVe=\Ve/m_p, \ \tVm=\Vm/m_p$. Mapping to our parameterization (\ref{GenParam}), we have 
\begin{equation}
    \begin{split}
        \delta_1=A, \ \ \ \ 
        \gamma_0=B, \ \ \ \ 
        \sigma_1=\frac{1}{2}\nu_1=C.
    \end{split}
\end{equation}
Following the discussion from Sec.~\ref{sec:genericlagrangian} in deriving Eq.~(\ref{eq:hrlgenmod}), we can write the solution for propagating GWs for this model as
\begin{align}
    h_{\text{R,L}} &= \Bar{h}_{\text{R,L}}\text{exp}\bigg[\frac{i(1+z)}{2}\bigg(\pm\gamma_0\frac{z_0}{1+z} + \delta_1kD_2 + \frac{\nu_1}{k}\int H dz + \frac{\sigma_1}{k}\int \frac{H_z}{1+z}dz\bigg)\bigg].
\end{align}
The GWs in this model satisfy the dispersion relation
\begin{align}
    \omega_{\text{R,L}}^2 = k^2\bigg(1 + \frac{\lambda_{\text{R,L}}}{k}\gamma_0\mathcal{H} + \delta_1 + \frac{\nu_1\mathcal{H}^2 + \sigma_1\mathcal{H}'}{k^2}\bigg),
\end{align}
and the group and phase velocities are
\begin{align}
    v_g^{\text{R,L}} &= 1 + \frac{1}{2}\delta_1 - \frac{\nu_1\mathcal{H}^2 + \sigma_1\mathcal{H}'}{2k^2}, \\
    v_p^{\text{R,L}} &= 1 +  \frac{1}{2}\delta_1+\frac{\lambda_{\text{R,L}}}{2k}\gamma_0\mathcal{H}  + \frac{\nu_1\mathcal{H}^2 + \sigma_1\mathcal{H}'}{2k^2}.
\end{align}

\subsection{Axion-dilaton-Chern-Simons-Gauss-Bonnet}\label{sec:axion-dilaton-chern-simons-gauss-bonnet}
Both the Chern-Simons and Gauss-Bonnet terms have been studied extensively in the literature (e.g.~\cite{Jackiw_2003, Alexander:2009tp, Yunes:2009hc, Sotiriou_2014}). Each term is a topological invariant; thus, they require a coupling to a pseudoscalar and scalar field respectively, which here we will call the axion and the dilaton. The Lagrangian is given by
\begin{equation}
    \mathcal{L}=-\frac{1}{2}(\partial\phi)^2-\frac{1}{2}(\partial\varphi)^2-\frac{\xi\phi}{\Lambda_E}\mathcal{G}-\frac{\tilde{\xi}\varphi}{\Lambda_O}R\tilde{R}, \label{eq:csgblagrangian}
\end{equation}
where $\mathcal{G}=R_{\mu\nu\rho\sigma}^2-4R_{\mu\nu}R^{\mu\nu}+R^2$ is the Gauss-Bonnet combination and $R\tilde{R}=R_{\mu\nu\rho\sigma}\tilde{R}^{\mu\nu\rho\sigma}$ is the Pontryagin density. In the notation of (\ref{genericLagrangian}), here we have $\zeta=\tilde{\zeta}=1,$ and 
\begin{equation}
    \mathcal{E}_{\mu\nu\rho\sigma}=\phi\Big(R_{\mu\nu\rho\sigma}-4g_{\mu\rho}R_{\nu\sigma}+g_{\mu\rho}g_{\nu\sigma}R\Big), \ \ \ \ \ \mathcal{O}_{\mu\nu\rho\sigma}=\varphi R_{\mu\nu\rho\sigma}
\end{equation}

\noindent
From Eq.~(\ref{eq:csgblagrangian}), we find the following field equations:
\begin{align}
&\Box\phi = -\frac{1}{\Lambda_E}\mathcal{G}, \\
&\Box\varphi = -\frac{1}{\Lambda_O}R\tilde{R}, \\
&G_{\mu\nu} + \frac{D_{\mu\nu}^{(\phi)}}{\Lambda_E} + \frac{2}{\Lambda_O}C_{\mu\nu} = 8\pi\bigg(T_{\mu\nu}^{(\phi)} + T_{\mu\nu}^{(\varphi)}\bigg),
\end{align}
where
\begin{align}
    D_{\mu\nu}^{(\phi)} &= (g_{\mu\rho}g_{\nu\sigma} + g_{\mu\sigma}g_{\nu\rho})\epsilon^{0\sigma\gamma\lambda}\nabla_{\kappa}[^{*}R^{\rho\kappa}_{~~\lambda\gamma}\phi'], \\
    C^{\mu\nu} &= (\nabla_{\alpha}\varphi)\epsilon^{\alpha\beta\gamma(\mu}\nabla_{\gamma}R^{\nu)}_{~\beta} + [\nabla_{(\alpha}\nabla_{\beta)}\varphi]^*R^{\beta(\mu\nu)\alpha}, \\
    T_{\mu\nu}^{(\phi)} &= \nabla_{\mu}\phi\nabla_{\nu}\phi - \frac{1}{2}g_{\mu\nu}\bigg(\nabla_{\alpha}\phi\nabla^{\alpha}\phi\bigg), \\
    T_{\mu\nu}^{(\varphi)} &= \nabla_{\mu}\varphi\nabla_{\nu}\varphi - \frac{1}{2}g_{\mu\nu}\bigg(\nabla_{\alpha}\varphi\nabla^{\alpha}\varphi\bigg).
\end{align}
Via Eqs.~(\ref{eq:hmunu}), (\ref{eq:graviton}) and (\ref{eq:flrw}), the propagation of GWs is given by the equation
\begin{align}
    h_{\text{R,L}}'' + (2\mathcal{H}+A)h_{\text{R,L}}' +k^2(1+B) h_{\text{R,L}} = 0, \label{eq:gwpropeom}
\end{align}
where 
\begin{align}
    A &= 16\lambda_{\text{R,L}}k\frac{\varphi'}{\Lambda_Oa}\mathcal{H} - \frac{16\phi'}{\Lambda_Ea}\mathcal{H}^2, \label{eq:bcsgb} \\
    B&=  \frac{4\phi'}{\Lambda_Ea}\mathcal{H} +\frac{\lambda_{\text{R,L}}k\varphi'}{\Lambda_Oa}\frac{24\mathcal{H}' - 8\mathcal{H}^2}{k^2}. \label{eq:ccsgb}
\end{align}
Mapping Eqs.~(\ref{eq:bcsgb}) and (\ref{eq:ccsgb}) to Eq.~(\ref{COsCEs}), we identify the parameters 
\begin{equation}\label{CSGBparams} \mu_0 = -16\phi', \ \ \  \alpha_1 = 16\varphi', \ \ \  \gamma_1 = 4\phi', \ \ \  \nu_2 = -8\varphi', \ \ \  \sigma_2 = 24\varphi', 
\end{equation}
with all other parameters being zero. Note that $m_p^{-2}$ multiplies all terms in (\ref{CSGBparams}), so that they are dimensionless as required.

Following the discussion again from Sec.~\ref{sec:genericlagrangian} in deriving Eq.~(\ref{eq:hrlgenmod}), for the right- and left-handed modes we have
\begin{align}
    h_{\text{R,L}} &= \bar{h}_{\text{R,L}}\text{exp}\bigg[\mp k(1+z)\bigg(\frac{8\varphi'_0}{\Lambda_O}z_1\bigg) + \frac{8\phi_0'}{\Lambda_E}\int Hdz\bigg] \nonumber \\ &\times \text{exp}\bigg\{i\bigg[\frac{k(1+z)\phi_0'}{\Lambda_E}z_1 \pm \frac{4\phi_0'}{\Lambda_Ok^2}\bigg(3\int (1+z)^{-1}H_zdz \mp \int Hdz\bigg)\bigg]\bigg\}. \label{eq:hrl-csgb}
\end{align}
GWs in this model satisfy the dispersion relation 
\begin{align}
    \omega_{\text{R,L}}^2 = k^2\bigg(1 + \frac{4\phi'}{\Lambda_Ea}\mathcal{H} +\frac{\lambda_{\text{R,L}}k\varphi'}{\Lambda_Oa}\frac{24\mathcal{H}' - 8\mathcal{H}^2}{k^2}\bigg), \label{eq:moddisp}
\end{align}
and the group and phase velocities are\footnote{It is worth noting that GW propagation for Palatini Chern-Simons gravity was computed in \cite{Sulantay2023} to quadratic order in the coupling constant. It was found that the amplitude birefringence is a more relevant effect than the velocity birefringence, which agrees with what we find in this work for the Chern-Simons contribution.} 
\begin{align}
    v_g^{\text{R,L}} &= 1 + \frac{2\phi'}{\Lambda_Ea}\mathcal{H}, \\
    v_p^{\text{R,L}} &= 1 + \frac{2\phi'}{\Lambda_Ea}\mathcal{H} + \frac{\lambda_{\text{R,L}}k\varphi'}{\Lambda_Oa}\frac{12\mathcal{H}' - 4\mathcal{H}^2}{k^2}.
\end{align}

\subsection{$U(1)$ vector fields}\label{sec:DarkU(1)}

Finally, we will illustrate the mapping with a gauge-invariant interaction Lagrangian where the Riemann tensor is coupled to the field strength of a (dark) photon, which are dimension-six operators. For simplicity, here we neglect dimension-four couplings of the form $\sim RA^2, \ R_{\mu\nu}A^\mu A^\nu$, which produce similar physics to the KR interactions of Sec.~\ref{sec:KRDM}. 

Since the photon field strength is also an antisymmetric rank-two tensor, we can include the same interactions as in the KR case. However, they are all suppressed by a cutoff:
\begin{equation}\label{LintMax}
\begin{split} 
\mathcal{L}_{int}&=\frac{\tilde{\xi}_1}{\Lambda_O^2}\tilde{R}^{\mu\nu\rho\sigma}F_{\mu\nu}F_{\rho\sigma}+\frac{1}{\Lambda_E^2}\Big(\xi_1 R^{\mu\nu\rho\sigma}F_{\mu\nu}F_{\rho\sigma}+\xi_2R^{\mu\rho}F_{\mu\lambda}F_{ \ \rho}^{ \lambda}+\xi_3RF_{\mu\nu}F^{\mu\nu}+\xi_4RF_{\mu\nu}\tilde{F}^{\mu\nu}\Big).
\end{split} 
\end{equation}
The gauge fields are described as usual by
\begin{equation}\label{VectorBackground}
    \mathcal{L}_{em}=-\frac{1}{4}F_{\mu\nu}F^{\mu\nu}-A_\mu J^\mu -V(A)
\end{equation}
for some source current $J^\mu$ and potential $V$.

The gravitational field equations will be analogous to each term in (\ref{KRgravEquations}) with $F_{\mu\nu}$ replacing $B_{\mu\nu}$, except we have the usual Maxwell energy momentum tensor replacing (\ref{KRTmunu}). Since the interactions in (\ref{LintMax}) contain terms that are essentially of the form $h\partial^2h (\partial A)^2,$ the field equations for $A_\mu$ will contain third derivative terms of the metric fluctuation in general. This is different from the KR model, where there were no derivatives of the matter fields in the non-minimal interactions. 

We will be assuming that the evolution of the vector fields is completely dominated by the background Lagrangian (\ref{VectorBackground}) and neglect the contributions from the interactions, but for completeness, the full field equations are 
\begin{equation}
\begin{split}
    \nabla_\alpha F^{\alpha\mu}&=J^\mu-\frac{\partial V}{\partial A_\mu}-\frac{2\tx_1}{\Lambda_O^2}\Big(2\tilde{R}^{\mu\lambda\alpha\beta}\nabla_\lambda F_{\alpha\beta}+F_{\alpha\beta}\nabla_\lambda\tilde{R}^{\mu\lambda\alpha\beta}+2F_{\alpha\beta}\varepsilon^{\mu\rho\sigma\lambda}\nabla_\lambda R^{\alpha\beta}_{ \ \ \ \rho\sigma}\Big) \\
    &+\frac{2}{\Lambda_E^2}\Big[2\xi_1\nabla_\lambda(R^{\mu\lambda\alpha\beta}F_{\alpha\beta})+\xi_2\Big(R^{\mu}_{ \ \beta}\nabla_\alpha F^{\alpha\beta}+\nabla_\alpha(R^\alpha_{ \ \beta}F^{\beta\mu})+\tfrac{1}{2}F^{\alpha\beta}\nabla_{[\alpha}R_{\beta]}^{ \ \mu}\Big) \\
    & \ \ \ \ \ +2\nabla_\alpha(\xi_3RF^{\mu\alpha}+\xi_4R\tilde{F}^{\mu\alpha})\Big].
\end{split}    
\end{equation}

Consider the scenario where the GWs pass through a region of slowly varying magnetic fields orthogonal to the propagation direction, in the $x-y$ plane. We parameterize the field strength tensor as
\begin{equation}
    F_{\mu\nu}=\frac{1}{\sqrt{2}}\begin{pmatrix}
        0&0&0&0\\ 
        0&0&0&-aB_y \\
        0&0&0&aB_x \\
        0&aB_y&-aB_x&0
    \end{pmatrix},
\end{equation}
where $B_x$ and $B_y$ are functions of $\eta$. Since $\vec{E}\cdot\vec{B}=0$ in this configuration, there is no contribution to the field equations from the $RF\tilde{F}$ interaction. 

\noindent
To lowest order in the magnetic fields, the GW equation of motion is 
\begin{equation}\label{heomwithmagfield}
    h_{R,L}''+(2\mathcal{H}+A)h_{R,L}'+k^2(1+B)h_{R,L}=0,
\end{equation}
with
\begin{equation}\label{origAB}
\begin{split}
    A&=-2\mathcal{H}\frac{(\xt-2\xth)}{(\Lambda_Ea)^2}\mathcal{B}^2+\frac{(\xt-2\xth)}{(\LE a)^2}\frac{d}{d\eta}\mathcal{B}^2, \\
    B&=-\frac{(6\xo-\xt)}{(\LE a)^2}\mathcal{B}^2+\frac{1}{k^2}\frac{4(\xo-\xt+3\xth)}{(\LE a)^2}\frac{d}{d\eta}(\mathcal{H}\mathcal{B}^2) -\lambda_{R,L}\frac{1}{k}\frac{\tx_1}{(\LO a)^2}\frac{d}{d\eta}\mathcal{B}^2,
\end{split}
\end{equation}
where $\mathcal{B}^2=(B_x^2+B_y^2)/m_p^2$. 

In assuming slowly-varying fields, we have neglected terms of the form $(d\mathcal{B}/d\eta)^2$ and $\tfrac{d^2}{d\eta^2}\mathcal{B}$. We see that the GWs are only sensitive to the magnitude of the magnetic fields in this configuration. The amplitude of the left- and right-handed strains are indeed modified, but in the same fashion, due to the dependence on the magnetic fields appearing with the same sign in the coefficient $A$. Thus, there is no amplitude birefringence in this model; however, the model predicts velocity birefringence from the last term in $B$. 

Although the interaction is dimension-six, we see the terms are appearing at low $k$-order. Moreover, the $\{\alpha,\beta,...\}$ coefficients in (\ref{COsCEs}) are dimensionless, while here, $\mathcal{B}$ is dimension-one. In order to map to (\ref{GenParam}), we  define the dimensionless quantity
\begin{equation}
    \tB=\frac{\mathcal{B}}{\Lambda a},
\end{equation}
which we assume to be nearly constant, and set a common cutoff scale\footnote{We can still accommodate the cutoff scales being significantly different orders of magnitude. Specifically, we define a new dimensionless constant $\tilde{\xi}_1'=\tx_1\frac{\Lambda_E}{\Lambda_O}.$ Assuming $\tilde{\xi}_1\sim O(1)$, the scale differences are characterized by $\tx_1'\ll 1$ if $\Lambda_E\ll\Lambda_O$ or $\tx_1'\gg 1$ if $\Lambda_E\gg\Lambda_O$. } $\Lambda_E=\Lambda_O\equiv \Lambda$. This implies that 
\begin{equation}
    \frac{1}{(\Lambda a)^2}\frac{d}{d\eta}\mathcal{B}^2=\frac{d}{d\eta}\tB^2+\mathcal{H}\tB^2\approx\mathcal{H}\tB^2.
\end{equation}
In terms of $\tB$, (\ref{origAB}) becomes
\begin{equation}
    \begin{split}
        A&=-(\xi_2-2\xi_3)\mathcal{H}\tB^2, \\
        B&=-(6\xi_1-\xi_2)\tB^2+\frac{1}{k^2}\Big(\xi_1-\xi_2+3\xi_3\Big)\Big[\mathcal{H}'+\mathcal{H}^2\Big]\tB^2-\lambda_{R,L}\frac{\tx_1}{k}\mathcal{H}\tB^2.
    \end{split}
\end{equation}
Mapping to (\ref{COsCEs}), we obtain
\begin{equation}\label{BfieldEvenParams}
    \begin{split}
        \alpha_0=-(\xi_2-2\xi_3)\tB^2, \ \ \ \delta_1=-(6\xi_1-\xi_2)\tB^2, \ \ \ \nu_1=\sigma_1=4(\xi_1-\xi_2+3\xi_3)\tB^2
    \end{split}
\end{equation}
for the parity-even contributions, and
\begin{equation}\label{BfieldOddParam}
    \gamma_0=-\tx_1\tB^2
\end{equation}
for the parity-odd contribution. The solution for the GWs is
\begin{align}
    h_{\text{R,L}} &= \Bar{h}_{\text{R,L}}\text{exp}\bigg[-\frac{1}{2}\alpha_0z_0+\frac{i(1+z)}{2}\bigg(\lambda_{R,L}\gamma_0\frac{z_0}{1+z} + \delta_1kD_2 + \frac{\nu_1}{k}\int H dz + \frac{\sigma_1}{k}\int \frac{H_z}{1+z}dz\bigg)\bigg].
\end{align}
while the dispersion relation, group, and phase velocities are 
\begin{align}
    \omega_{\text{R,L}}^2 = k^2\bigg(1 + \frac{\lambda_{\text{R,L}}}{k}\gamma_0\mathcal{H} + \delta_1 + \frac{\nu_1\mathcal{H}^2 + \sigma_1\mathcal{H}'}{k^2}\bigg),
\end{align}
\begin{align}
    v_g^{\text{R,L}} &= 1 + \frac{1}{2}\delta_1 - \frac{\nu_1\mathcal{H}^2 + \sigma_1\mathcal{H}'}{2k^2}, \\
    v_p^{\text{R,L}} &= 1 +  \frac{1}{2}\delta_1+\frac{\lambda_{\text{R,L}}}{2k}\gamma_0\mathcal{H}  + \frac{\nu_1\mathcal{H}^2 + \sigma_1\mathcal{H}'}{2k^2}.
\end{align}

The effects coming from this model are likely never to be observed in any GW detector. If we assume $\Lambda\sim m_p,$ then the beyond-GR modification is quartic Planck suppressed, $\tB^2\sim \frac{B_x^2+B_y^2}{m_p^4}$. Nonetheless, the model illustrates the utility of parameterization presented in this work.

\subsection{Summary and numerical illustration}\label{sec:SummaryAndPlots}
\begin{figure*}[htb!]
    \includegraphics[width=0.482\textwidth]{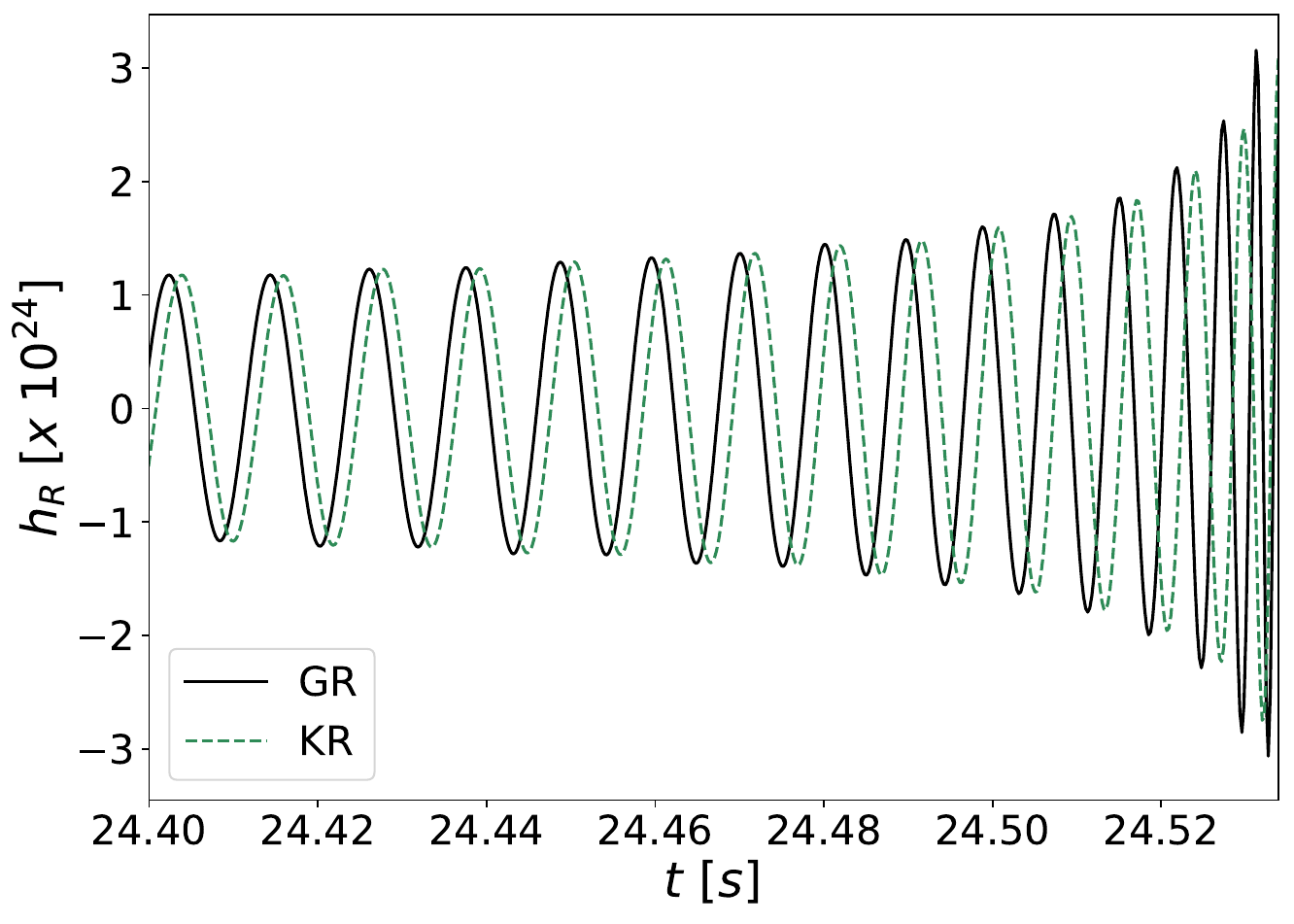}\hfill
    \includegraphics[width=0.5\textwidth]{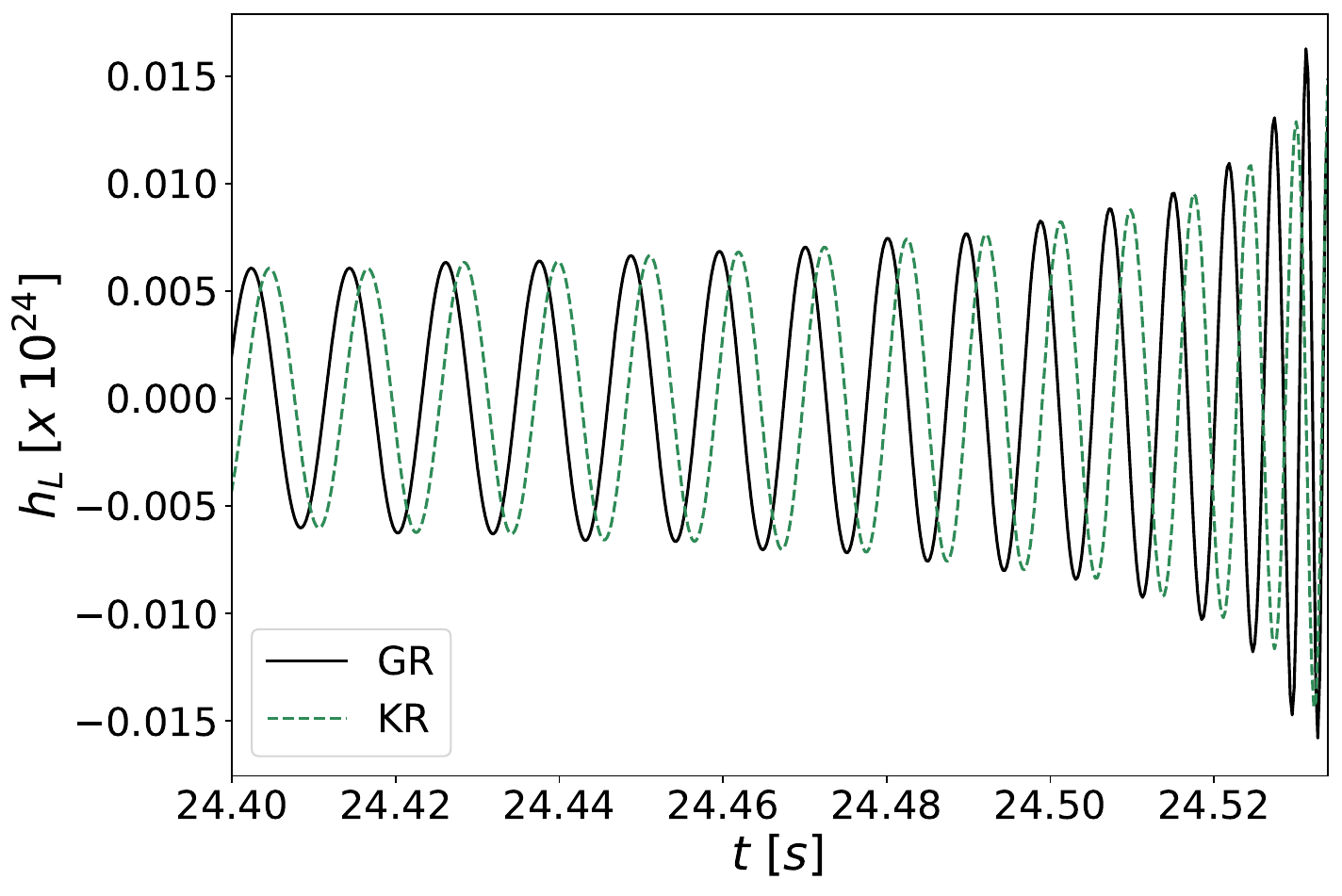}
    \caption{Example modification to a binary black hole waveform for $h_R$ (left) and $h_L$ (right) in the KR model. To generate the waveform, we employ the \texttt{GW Analysis Tools} code \cite{Perkins:2021mhb}. We see that the amplitude is slightly suppressed due to the parity-invariant modification to the amplitude, while there is a larger phase shift in the left-handed mode compared to the right-handed mode, due to the parity-violating phase modification. For the source parameters we take $m_1 = 20 M_\odot$, $m_2 = 18 M_\odot$, $\iota = 2.6$ rad, $\psi = 3.14$ rad, $RA = 3.45$ rad, $Dec = -3968$ rad. For computational ease we rescale $f/100$ Hz, and $D_2/\text{Gpc}$. The modification parameters are chosen to be artificially large in order to visually see the effects; in dimensionless units $\tilde{\mathcal{V}}_e = \tilde{\mathcal{V}}_m = 3$.}
    \label{fig:waveform-kr}
\end{figure*}

\begin{figure*}[htb!]
    \includegraphics[width=0.482\textwidth]{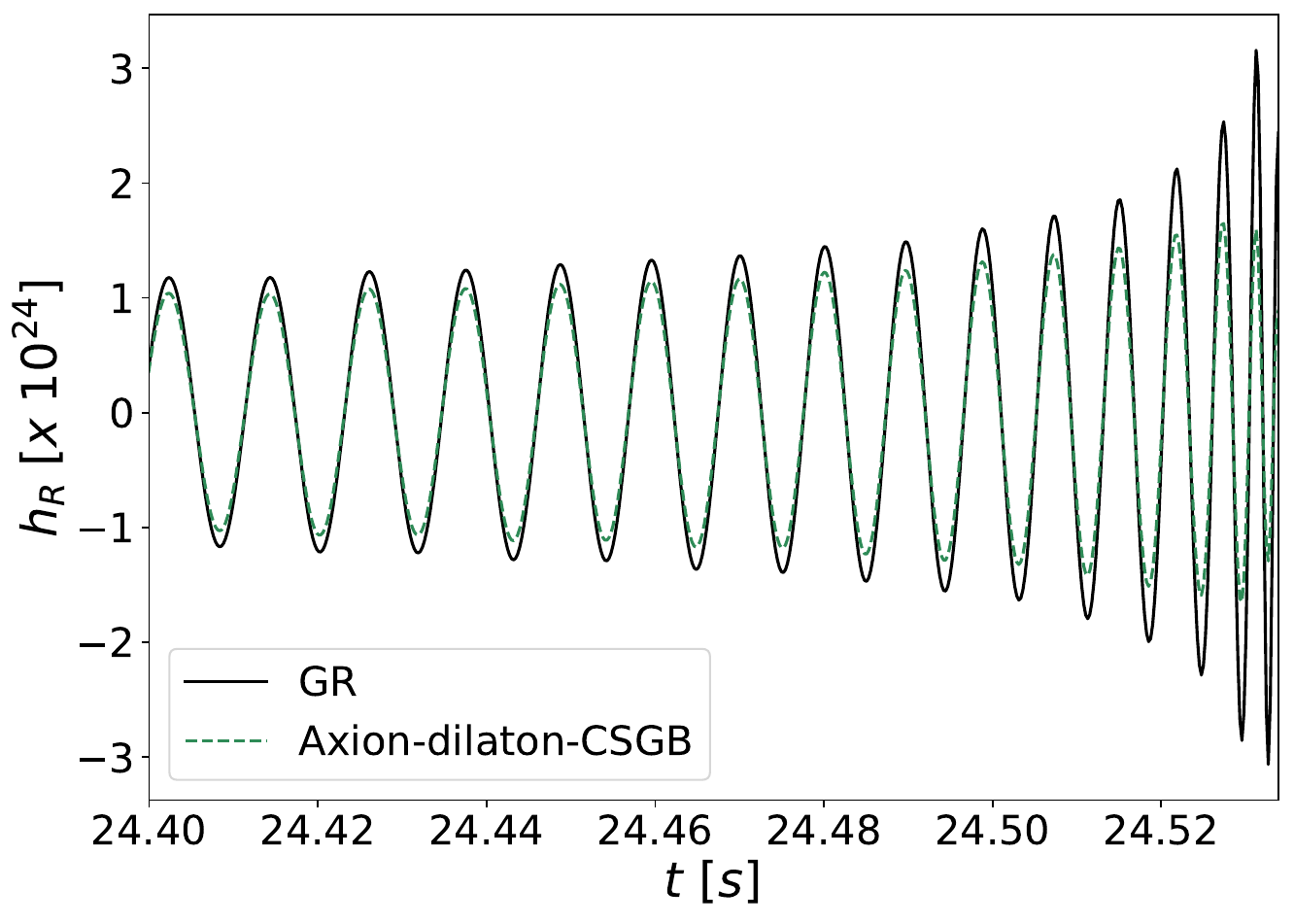}\hfill
    \includegraphics[width=0.5\textwidth]{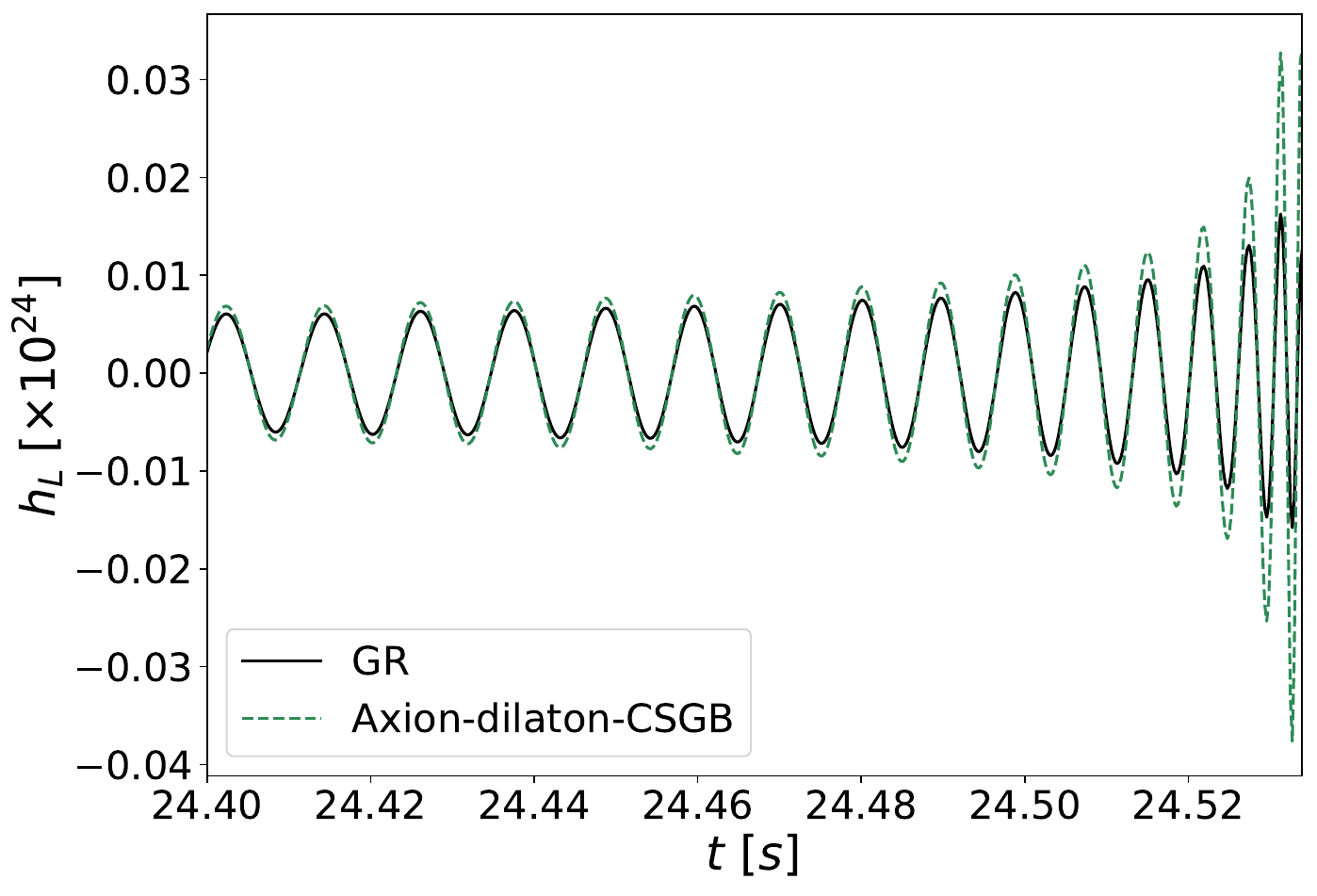}
    \caption{Example modification to a binary black hole waveform for $h_R$ (left) and $h_L$ (right) in the axion-dilaton-Chern-Simons-Gauss-Bonnet model. The right-handed mode is attenuated while the left-handed mode is amplified, due to the parity-violating modification to the amplitude. The phase shift is a much smaller effect, and it arises due to the parity-invariant phase modification. Other notes from Fig.~\ref{fig:waveform-kr} apply here, with the modification parameters again being chosen to be artificially large in order to visually see the effects; in dimensionless units $\phi_0' = \varphi_0' = 6$.}
    \label{fig:waveform-axidil}
\end{figure*}

\begin{figure*}[htb!]
    \includegraphics[width=0.482\textwidth]{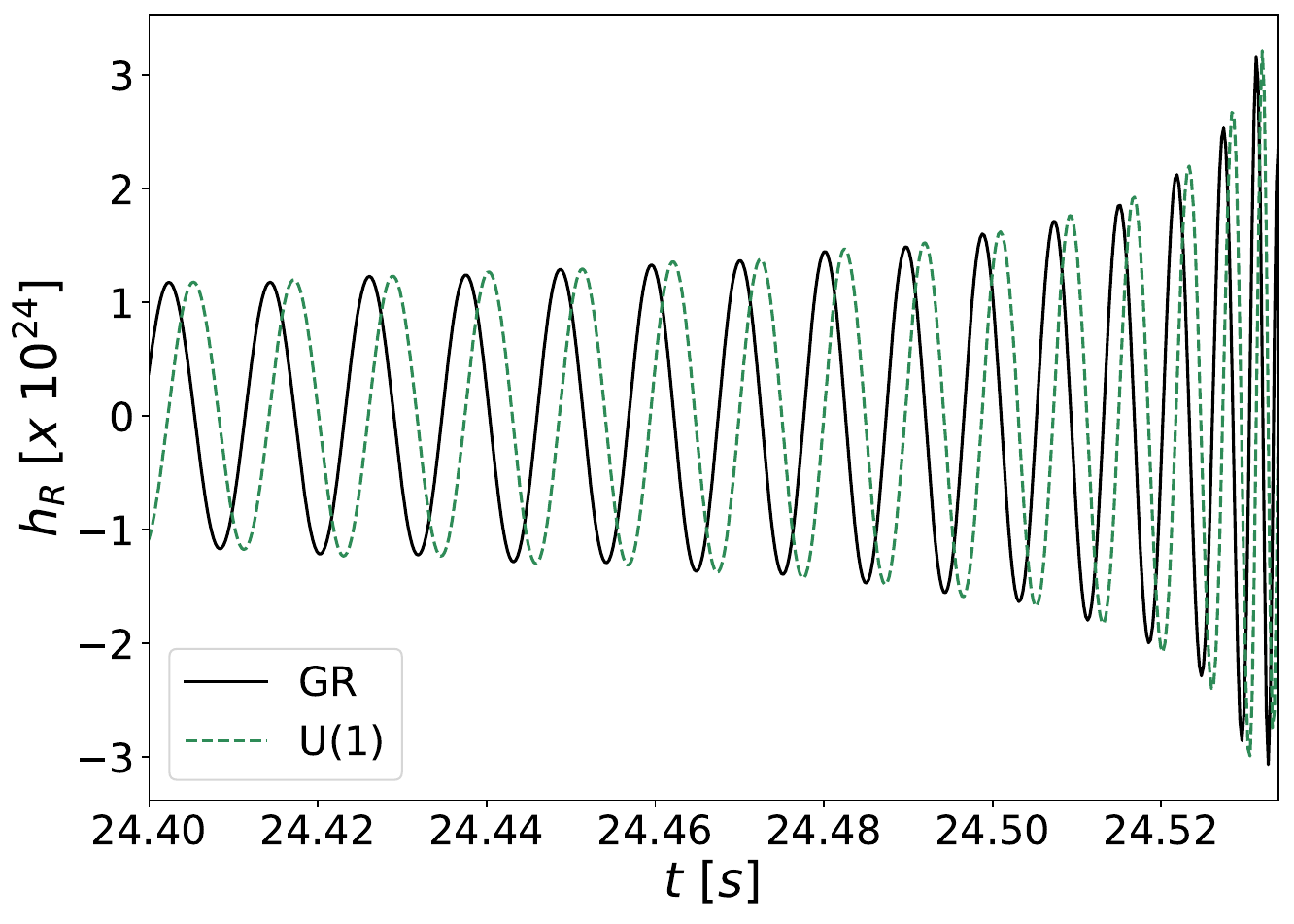}\hfill
    \includegraphics[width=0.5\textwidth]{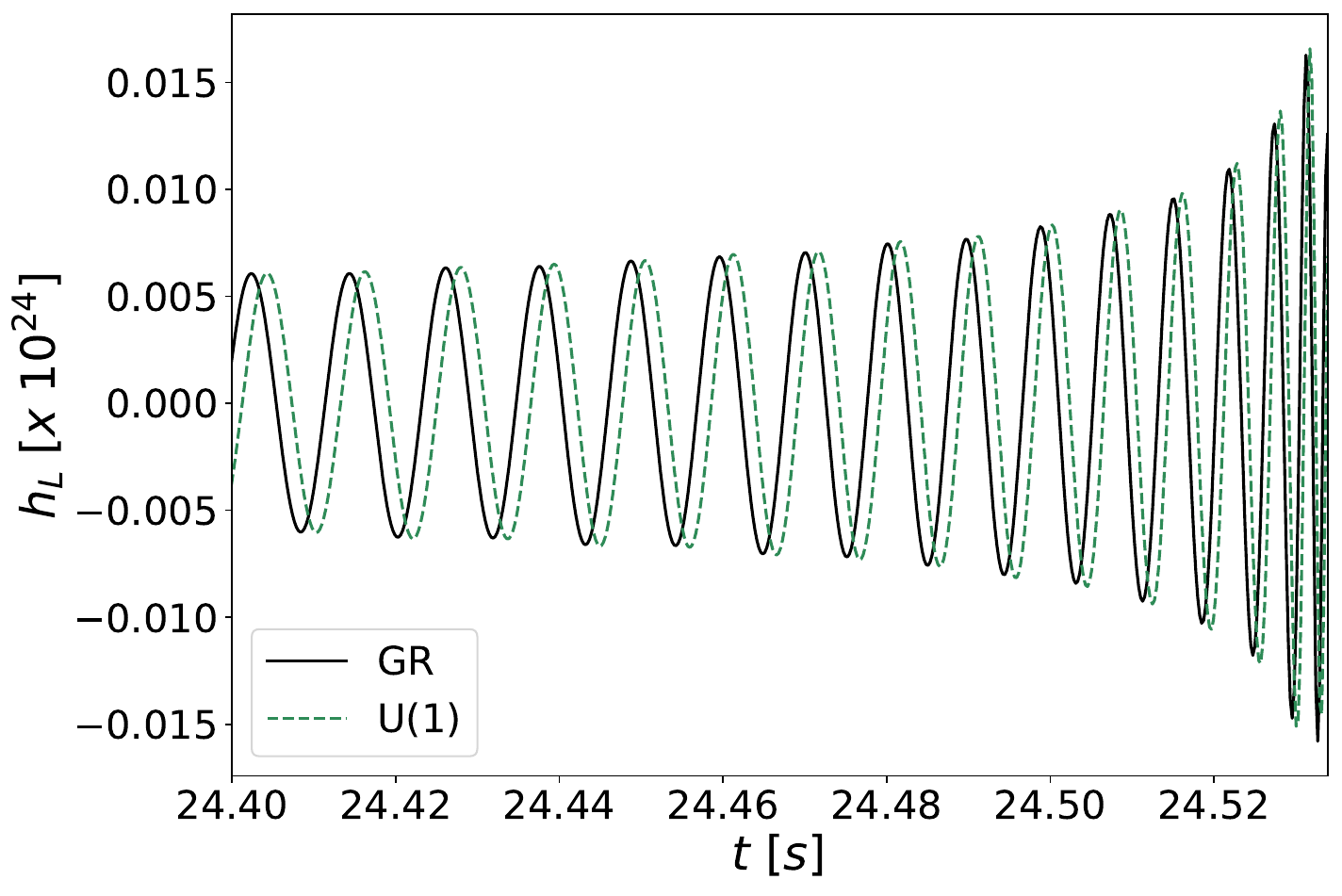}
    \caption{Example modification to a binary black hole waveform for $h_R$ (left) and $h_L$ (right) in the $U(1)$ model. The phase shift affects the right-handed mode more than the left-handed mode, due to the parity-violating modification. The amplitude modification is a much smaller effect, and it is due to the parity-invariant modification to the amplitude. Other notes from Fig.~\ref{fig:waveform-kr} apply here, with the modification parameters again being chosen to be artificially large in order to visually see the effects; in dimensionless units $\tilde{B}^2 = 2$.}
    \label{fig:waveform-U1}
\end{figure*}

The three theories we have analyzed each contribute multiple beyond-GR parameters or functions, with the KR and dark photon models being the most similar. We summarize all of the non-zero terms in Table \ref{table1} and illustrate the effects of each model on the waveform in Figs.~\ref{fig:waveform-kr}-\ref{fig:waveform-U1}. For the figures, we use a frequency-domain inspiral waveform in the stationary phase approximation, which is effectively a TaylorF2-type inspiral waveform, written as $\tilde{h}(f) = \mathcal{A}(f)e^{i\Psi(f)}$ \cite{Sathyaprakash1991,Droz1999,Buonanno2009}. We inherit the standard GR inspiral structure, in which the phase is computed up to the usual PN expansion (3.5PN TaylorF2-type phasing, as commonly used in LIGO analyses \cite{Blanchet2002,Arun:2008kb,LIGOScientific:2018mvr}), while the amplitude is kept at leading PN order (``restricted" waveform) \cite{Cutler1994}. The parity-violating corrections are implemented only at the level of propagation, where amplitude birefringence enters at relative 1.5PN order, while phase birefringence enters at a much higher PN order ($\sim$5.5 PN); this is consistent with the frequency-dependent corrections expected from modified dispersion relations and ppE-type parameterizations \cite{Yunes:2009hc,Mirshekari_2012}.

\begin{table}[h]\label{table1}
\centering
\begin{tabular}{lcccccccccc}
&$\alpha_0$ &$\alpha_1$& $\gamma_0$ & $\gamma_1$ & $\delta_1$ & $\mu_0$ & $\sigma_1$ & $\sigma_2$ & $\nu_1$ & $\nu_2$\\ \hline
Kalb-Ramond &  &  & \checkmark  & & \checkmark & & \checkmark & & \checkmark &  \\
Chern-Simons-Gauss-Bonnet &  & \checkmark  &  & \checkmark & & \checkmark & & \checkmark & & \checkmark \\ 
 Dark photon & \checkmark & & \checkmark & & \checkmark & & \checkmark & & \checkmark & \\ \hline
\end{tabular}
\caption{Summary of all non-zero terms in the three models under consideration.}
\end{table}

\section{Observational Constraints}\label{sec:constraints}
\subsection{Constraints from GW170817}
In this section, we illustrate the ease with which the parameterization (\ref{GenParam}) and (\ref{COsCEs}) can be used to constrain model parameters. 
The coincident gamma ray burst (GRB)/GW signal GW170817 \cite{LIGOScientific:2017zic}  gives one of the most remarkable constraints on beyond-GR models to date. Namely, the time delay between the observed GRB and GW gives a constraint on the propagation speed of GWs:
\begin{equation}\label{cgwConstraint}
    -3\times 10^{-15}<c_{\text{gw}}-1<7\times 10^{-16}.
\end{equation}
This directly translates to a constraint on the group velocity, (\ref{gengroupvelocity})\footnote{It is possible for the propagation time difference between light and GWs to accumulate in highly dense DM environments. This phenomenon can result in a relaxed constraint on $c_{\text{gw}}$, as was shown in \cite{Zhang:2023fhs}.}. 

In \cite{Jenks:2023pmk}, the authors were studying parity-odd terms and focused on constraining the $\delta$-functions, which are the lowest order contributions to the group velocity for late times. This is of particular interest for  ghost-free scalar-tensor theory, symmetric teleparallel equivalent of GR and Horava-Lifshitz
theory, as $\delta_2\neq 0$ for each. In our case, we instead have only parity-even terms: $\{\delta_1,\nu_1,\sigma_1\}$ for the KR model and the dark photon model, and $\gamma_1$ for axion-dilaton-Chern-Simons-Gauss-Bonnet:
\begin{equation}
    \begin{split}
        v_g&=1+\frac{1}{2}\Big(\delta_1-\frac{\nu_1\mathcal{H}^2+\sigma_1\mathcal{H}'}{k^2}\Big), \\
        v_g&=1+\frac{\gamma_1\mathcal{H}}{2\Lambda_Ea}.
    \end{split}
\end{equation}
First for the KR and dark photon models, focusing on late times, we constrain $\delta_1$ to be 
\begin{equation}\label{delta1constraint}
    |\delta_1|<6\times 10^{-15},
\end{equation}
using the weaker of the bounds in (\ref{cgwConstraint}). The contributions to $\delta_1$ in both models come from the $\xi_1$ contraction with the Riemann tensor and the $\xi_2$ with the Ricci tensor. 

\noindent
For the KR model, $\delta_1=(\tVm^2+\tVe^2)(\xi_2-6\xi_1),$ corresponding to 
\begin{equation}
    \frac{\sqrt{\Vm^2+\Ve^2}}{m_p}\sqrt{|6\xi_1-\xi_2|}<10^{-8}.
\end{equation}
The quantity $\sqrt{\Vm^2+\Ve^2}$ essentially represents the energy scale of the KR field, which we can take to be the order of its mass, $m_{\text{B}}$. The dimensionless constants $\xi_i$ are assumed to be $O(1)$, so taking $\sqrt{|6\xi_1-\xi_2|}\sim 1$, we arrive at 
\begin{equation}\label{mbconstraint}
    m_{\text{B}}\lesssim 10^{11} \text{ GeV},
\end{equation}
using $m_p\sim 10^{19}$ GeV. If the KR fields were to constitute a significant portion of the relic DM density, the upper limit on their mass is closer to $10^6$ eV, depending on the production mechanism \cite{Capanelli:2023uwv}. The constraint (\ref{mbconstraint}) is clearly many orders of magnitude away from that limit, although this simplified scenario of the fields in a static configuration will need to be relaxed in order to consider a more physically realistic DM - black hole system.

As mentioned in Sec. \ref{sec:DarkU(1)}, detecting the effects of the non-minimal couplings for the dimension six operators of the $U(1)$ model is extremely unlikely due to the Planck suppression. For completeness, using $\delta_1$ from (\ref{BfieldEvenParams}) along with (\ref{delta1constraint}), we find $\frac{|B|^2}{\Lambda^2}< 10^{23} \text{ (GeV)}^2$, again assuming $\sqrt{|6\xi_1-\xi_2|}\sim 1$. 

Lastly for Chern-Simons-Gauss-Bonnet, we have the constraint on $\gamma_1$ and the cutoff scale as
\begin{equation}
    \Big|\frac{\gamma_1\mathcal{H}}{\Lambda_Ea}\Big|=\Big|\frac{\gamma_1H}{\Lambda_E}\Big|<6\times 10^{-15}, \label{eq:csgb-constraint}
\end{equation}
where $H=\mathcal{H}/a$ is again the Hubble parameter in cosmic time. Then from (\ref{CSGBparams}), we have that $\gamma_1=4\phi'/m_p^2$, where $\phi'$ is the derivative with respect to conformal time. Thus, we can evaluate Eq.~(\ref{eq:csgb-constraint}) at different epochs in the universe. 

Since the upper bound of $\phi' \lesssim 10^{-22}$ eV$^2$ was obtained in \cite{Daniel:2024lev} for Chern-Simons-Gauss-Bonnet gravity in the late universe, we will focus here on the upper bound for the very early universe. We see that the further back in time we go, the tighter the upper bound on $\phi'$ from Eq.~(\ref{eq:csgb-constraint}), since $\phi'$ and $H$ both appear in the numerator of that equation. The most stringent constraint we obtain on $\phi'$ is thus at the highest value of $H$ at which Eq.~(\ref{eq:csgb-constraint}) is still valid, which is at the boundary of the EFT validity ($H \sim \Lambda_E$):

\begin{equation}
    |\phi'| \lesssim 10^{-15}m_p^2. \label{eq:phi-prime-csgb-constraint}
\end{equation} 
While the Chern-Simons-Gauss-Bonnet modifications are constrained to be extremely small in the late universe to reproduce observational data, it is ongoing work to investigate whether the theory reproduces the standard cosmological paradigms at early times, as a follow up to \cite{Ortega:2024prv} by the same authors. Therefore, one can expect that the Chern-Simons-Gauss-Bonnet modifications may be significantly larger in the early universe. Eq.~(\ref{eq:phi-prime-csgb-constraint}) gives us an indication of how large some of these modifications can be in the regime of validity for the EFT, relative to the Planck scale.

\subsection{Constraints on gravitational parity and Lorentz violation using ground- and space-based detectors}
To further illustrate the utility of our parameterization, we map our framework to the constraints in \cite{Zhang:2025kcw}, where GW propagation is parameterized as
\begin{align}
    h_A'' + (2 + \bar{\nu} + \nu_A)\mathcal{H}h_A' + (1 + \bar{\mu} + \mu_A)k^2h_A = 0.
\end{align}
Here $\mu$ and $\nu$ are frequency dependent and encode parity violation, while $\bar{\mu}$ and $\bar{\nu}$ are frequency independent and encode Lorentz violation; $A = L,R$ denotes the helicity mode. $\mu$, $\nu$, $\bar{\mu}$ and $\bar{\nu}$ are given by the following equations:
\begin{align}
    \mu_A &= \rho_A\alpha_{\mu}(\tau)\bigg(\frac{k}{aM_{PV}}\bigg)^{\beta_{\mu}}, \label{eq:mu_A} \\
    \mathcal{H}\nu_A &= \bigg[\rho_A\alpha_{\nu}(\tau)\bigg(\frac{k}{aM_{PV}}\bigg)^{\beta_{\nu}}\bigg]', \label{eq:Hnu_A}
\end{align}
for the parity-violating terms, and
\begin{align}
    \bar{\mu} &= \alpha_{\bar{\mu}}(\tau)\bigg(\frac{k}{aM_{LV}}\bigg)^{\beta_{\bar{\mu}}}, \\
    \mathcal{H}\bar{\nu} &= \bigg[\alpha_{\bar{\nu}}(\tau)\bigg(\frac{k}{aM_{LV}}\bigg)^{\beta_{\bar{\nu}}}\bigg]' \label{eq:Hnubar}
\end{align}
for the Lorentz-violating terms. In Eqs.~(\ref{eq:mu_A})-(\ref{eq:Hnubar}), $\rho_A \equiv \pm 1$ for $A = R,L$ respectively, $\alpha_{\mu}, \alpha_{\nu}, \alpha_{\bar{\mu}}$ and $\alpha_{\bar{\nu}}$ are dimensionless functions of conformal time $\tau$, $\beta_{\mu}$, $\beta_{\nu}$, $\beta_{\bar{\mu}}$ and $\beta_{\bar{\nu}}$ are numbers, $M_{PV}$ and $M_{LV}$ are the energy scales of parity and Lorentz violation, respectively, and primes denote conformal time derivatives.

We seek to map our framework to \cite{Zhang:2025kcw}. We start by looking at the $n=0$ term of our $C_O^{(1)}$ in Eq.~(\ref{COsCEs}), which reads
\begin{align}
    C_O^{(1)} = \lambda_{\text{R,L}}\bigg(\frac{k\alpha_1}{\Lambda_Oa}\mathcal{H} + \beta_1\bigg). \label{eq:CO1n0}
\end{align}
We can map this to the $\beta_{\nu} = 1$ term in Eq.~(\ref{eq:Hnu_A}):
\begin{align}
    \mathcal{H}\nu_A({\beta_{\nu} = 1})= \bigg(\rho_A\alpha_{\nu}\frac{k}{aM_{PV}}\bigg)' = \frac{\rho_Ak}{aM_{PV}}(\alpha_{\nu}'- \mathcal{H}\alpha_{\nu})h_A'. \label{eq:HnuAbetanu1}
\end{align}
Comparing Eqs.~(\ref{eq:CO1n0}) and (\ref{eq:HnuAbetanu1}), we can make the identification that $\alpha_{\nu}({\beta_{\nu}} = 1) = -\alpha_1$.

\noindent
Next, we look at the $\beta_{\mu} = 1$ term in Eq.~(\ref{eq:mu_A}):
\begin{align}
    \mu_A = \rho_A\alpha_{\mu}\frac{k}{aM_{PV}}.
\end{align}
Upon comparing it to the $m=1$ term of the $C_O^{(0)}$ coefficient in Eq.~(\ref{COsCEs}),
\begin{align}
    C_O^{(0)} = \lambda_{\text{R,L}}k\bigg[\frac{\gamma_2}{(\Lambda_Oa)^2}\mathcal{H} + \frac{\delta_2}{\Lambda_Oa}\bigg],
\end{align}
we can identify that $\alpha_{\mu}(\beta_{\mu} = 1) = \delta_2$.

Comparing different values of $m$ and $n$ to different numbers for $\beta_{\mu}$, $\beta_{\nu}$, $\beta_{\bar{\mu}}$ and $\beta_{\bar{\nu}}$, we find the following mapping:
\begin{align}
    \alpha_{\mu,\bar{\mu}}(\beta_{\mu,\bar{\mu}}=m-1) &= \delta_m, \\
    \alpha_{\nu,\bar{\nu}}(\beta_{\nu,\bar{\nu}}=n) &= -\frac{1}{n}\alpha_n.
\end{align}
For the various values of $\beta_{\mu}$, $\beta_{\nu}$, $\beta_{\bar{\mu}}$ and $\beta_{\bar{\nu}}$ that are constrained in \cite{Zhang:2025kcw}, only the $\beta_{\nu} = 1$ constraint maps to a nonzero parameter for the three models we have considered ($\alpha_1$ for Chern-Simons-Gauss-Bonnet). For a given constraint $M_{PV} \gtrsim M_0$, $\alpha_{\nu} \lesssim 1/M_0$; we can then map the numerical $M_{PV}$ constraint to the constraint on $\varphi'$ in Chern-Simons-Gauss-Bonnet via Eq.~(\ref{CSGBparams}) after multiplying by $\mathcal{H}$ and a cutoff scale $\Lambda$ to account for natural units \cite{Jenks:2023pmk}.

We find the most stringent constraint on $\varphi'$ to come from the ground-based constraints, in particular GW190521ET+CE; the space-based constraints place a smaller lower bound on $M_{PV}$, which translates to a higher upper bound on $\varphi'$. The GW190521ET+CE constraint translates to an upper bound today of 
\begin{align}
    \varphi' \lesssim 10^{-21}~\text{eV}^2, \label{eq:varphi-constraint-today}
\end{align}
where we have evaluated $\mathcal{H} \equiv aH$ today using $a_0 = 1, H_0 \sim 10^{-33}~\text{eV}$, and we set $\Lambda = 100~\text{eV}$, which saturates the lower bound in \cite{Jenks:2023pmk}.

We observe that Eq.~(\ref{eq:varphi-constraint-today}) is a more stringent upper bound for the parity-violating sector of Chern-Simons-Gauss-Bonnet compared to $\varphi' \lesssim 10^{-15}~\text{eV}^2$ in \cite{Daniel:2024lev}, although here we are not considering the axion and dilaton to be kinetically coupled. Nevertheless, by considering different models outside of what was studied in this paper, it may be possible to constrain the theory parameters using both our parameterization and the framework in \cite{Zhang:2025kcw}. This is a direction for future work.

\section{Conclusion}\label{sec:conclusion}

In this work, we have presented a general parameterization for GW amplitudes experiencing non-minimal couplings to (dark) matter, capturing parity-even and parity-odd effects. The parameterization extends what was studied in \cite{Jenks:2023pmk,Daniel:2024lev} to account for departures from GR at orders $\mathcal{H}'/k^2$ and $\mathcal{H}^2/k^2$. We then studied three models where the interactions were dimension four, five, and six respectively, and mapped the fields to the general parameterization. The departure from GR for a late-time BH-BH merger is depicted in Figures \ref{fig:waveform-kr}, \ref{fig:waveform-axidil}, and \ref{fig:waveform-U1}, where we stress that the parameter values used for the plots are unphysically large  and are meant for illustrative purposes. For parity-even operators, gravitational waveforms exhibit amplitude and/or phase modifications independent of $k$ and at $O(k^{-2})$, which appear with the same sign for both left- and right-handed polarizations. The parity-odd operators modify the amplitude and/or phase with even powers of $k$, and contribute opposite signs for the left- and right-handed polarizations, predicting birefringence. 

We focused on three models where the non-minimal couplings were either linear or quadratic in the spacetime curvature. For these models, the dispersion relation (\ref{gendispersion}) was modified at order $k^0, \ k^{-1}$, and $k^{-2}$. Our parameterization is in principle applicable for modifications at arbitrarily large, positive powers of momentum, which are expected from higher curvature theories such as cubic gravity and higher derivative theories. This is easy to see, as equations of motion for such theories contain terms $\sim\partial^4h\sim k^4h$, potentially producing non-vanishing $\gamma_3,\delta_3$ in (\ref{COsCEs}), for example. It would be interesting to explore classes of models that exhibit departures from GR that are relevant primarily for high frequency GWs, and the ways in which they map to our parameterization.

The couplings to the Riemann curvature and its contractions, as written in (\ref{genericLagrangian}), do not encompass all types of non-minimal couplings to gravity. In particular, our parameterization (\ref{GenParam}) and (\ref{COsCEs}) potentially does not capture GW modifications that come from generalizations of teleparallel gravity \cite{Cai:2015emx,Bahamonde:2021gfp}, which may assume a vanishing Riemann tensor. An interesting example of a parity-violating operator in generalized teleparallel gravity is the Nieh-Yan term \cite{NIEH1981113,Li:2020xjt}, which has been studied in the context of the stochastic GW background in \cite{Cai:2021uup} and more recently in \cite{Xu:2024kwy}. This term is mass dimension three, and a parity-violating correction appears in the equation of motion for the GW amplitudes as $m_p^2\Box h_{L,R}\pm \tilde{m}\varphi'kh_{L,R}=0$, independent of the Hubble parameter (here $\varphi$ is an axion and $\tilde{m}$ has dimensions of mass). This term can not be mapped to (\ref{COsCEs}), and therefore an even more general, model-independent parameterization of the GW equations of motion would be needed to encompass all beyond-GR theories. 

Our approach to understanding the generic behavior of the GW amplitudes relied on a simplified description of the matter field configurations, as well as the assumption that they are slowly varying. The latter allowed us to write down a very general solution, (\ref{eq:hrlgenmod}), derived below in Appendix \ref{sec:AppendixA:GenProp}\footnote{We note that although the solution assumes slowly-varying matter fields, the modified dispersion relation $\omega(k)$ can be read off of the equations of motion for the GW amplitudes, and therefore does not rely on the same assumption.}. However, the $\mathcal{H}^2/k^2$ and $\mathcal{H}'/k^2$ corrections in our parameterization have importance in the early universe, when matter fields are not necessarily slowly varying. The equations of motion for the GW amplitudes in this scenario would need to be integrated numerically, which may be manageable if we neglect backreaction and assume the matter fields satisfy free equations of motion. The couplings will affect the primordial GW power spectrum and in turn the tensor-to-scalar ratio $r$, which is expected to be probed to $\mathcal{O}(10^{-3})$ in future experiments such as LiteBIRD \cite{LiteBIRD:2022cnt} and the Simons Observatory \cite{SimonsObservatory:2018koc}. It may therefore be possible to use constraints on $r$ to indirectly constrain the models under consideration. We leave a further in-depth analysis of the early-universe cosmology of these types of non-minimally coupled models for future work.

\subsection*{Acknowledgments}

 The authors thank Robert Brandenberger, Leah Jenks, Antonino Marciano and Hong-Yi Zhang for useful discussions and helpful comments on a draft of this work. S.A., T.D. and T.M. acknowledge support from the Simons Foundation, Award 896696. T.D. is supported in part by a Trottier Space Institute fellowship and by funds from NSERC.

\appendix
\section{General solution for slowly varying fields}\label{sec:AppendixA:GenProp}
Here we provide the derivation of Eq.~(\ref{eq:hrlgenmod}), the general parameterization of the corrections to $h_{\text{R,L}}$. The linear perturbations of GWs can be expressed in spatial Fourier space as
\begin{align}
    h_{\text{R,L}}(\eta,x) = \mathcal{A}_{\text{R,L}}(\eta)e^{-i[\theta(\eta) - k_ix^i]}. \label{eq:hfourier}
\end{align}
Plugging Eq.~(\ref{eq:hfourier}) into the equations of motion Eq.~(\ref{GenParam}), we find that the effective modified dispersion relation can be parameterized as
\begin{align}
    i\theta'' &+ \theta'^2 + i\theta'\{2\mathcal{H} + C_O^{(1)} + C_E^{(1)}\}  - k^2\{1 + C_O^{(0)} + C_E^{(0)}\} = 0, \label{eq:moddispgen}
\end{align}
with $C_O^{(1)}, C_E^{(1)}, C_O^{(0)}$ and $C_E^{(0)}$ given by Eq.~(\ref{COsCEs}). From here, we can linearize the equations of motion by taking $\theta \rightarrow \bar{\theta} + \delta\theta$, where the background $\theta$ is the usual GR solution, $\theta' = k - i\mathcal{H}$. Applying this to Eq.~(\ref{eq:moddispgen}), and performing a series expansion assuming that $\delta\theta \ll \bar{\theta}$, $\theta'' \ll (\theta')^2$ and $\delta\theta'' \ll \bar{\theta}\delta\theta'$, we get that
\begin{align}
    \delta\theta &= -i(\lambda_{\text{R,L}})^{2n}(\delta\theta_{E,A} + \lambda_{\text{R,L}}\delta\theta_{O,A})+ (\lambda_{\text{R,L}})^{2m+1}(\delta\theta_{O,V} + \lambda_{\text{R,L}}\delta\theta_{E,V}),
\end{align}
where the even and odd components of the amplitude and velocity birefrigence contributions are
\begin{align}
    \delta\theta_{E,A}' &= \frac{k^{2n}}{2}\bigg[\frac{\alpha_{2n}}{(\Lambda_Ea)^{2n}}\mathcal{H} + \frac{\beta_{2n}}{(\Lambda_Ea)^{2n-1}} + \frac{\mu_{2n}\mathcal{H}^2 + \rho_{2n}\mathcal{H}'}{(\Lambda_Ea)^{2n+1}}\bigg], \label{eq:deltathetaeaprime} \\
    \delta\theta_{O,A}' &= \frac{k^{2n+1}}{2}\bigg[\frac{\alpha_{2n+1}}{(\Lambda_Oa)^{2n+1}}\mathcal{H} + \frac{\beta_{2n+1}}{(\Lambda_Oa)^{2n}} + \frac{\mu_{2n+1}\mathcal{H}^2 + \rho_{2n+1}\mathcal{H}'}{(\Lambda_Oa)^{2n+2}}\bigg], \label{eq:deltathetaoaprime} \\
    \delta\theta_{O,V}' &= \frac{k^{2m}}{2}\bigg[\frac{\gamma_{2m}}{(\Lambda_Oa)^{2m}}\mathcal{H} + \frac{1}{\Lambda_Oa)^{2m-1}}\bigg(\delta_{2m} + \frac{\nu_{2m}\mathcal{H}^2 + \sigma_{2m}\mathcal{H}'}{k^2}\bigg)\bigg], \label{eq:deltathetaovprime} \\
    \delta\theta_{E,V}' &= \frac{k^{2m+1}}{2}\bigg[\frac{\gamma_{2m+1}}{(\Lambda_Ea)^{2m+1}}\mathcal{H} + \frac{1}{(\Lambda_Ea)^{2m}}\bigg(\delta_{2m+1} + \frac{\nu_{2m+1}\mathcal{H}^2 + \sigma_{2m+1}\mathcal{H}'}{k^2}\bigg)\bigg]. \label{eq:deltathetaevprime}
\end{align}
To simplify Eqs.~(\ref{eq:deltathetaeaprime})-(\ref{eq:deltathetaevprime}) further, we will assume that $\phi'$ and $\varphi'$ vary slowly with respect to the expansion of the universe, and therefore they can be well approximated by their current values via a Taylor expansion, e.g. $\phi' \approx \phi'_0$. Furthermore, we will use the fact that $dt = -dz/[H(z)(1+z)]$, as well as the fact that $k$ is a constant in conformal time. Using all of this, we can write the integrals of Eqs.~(\ref{eq:deltathetaeaprime})-(\ref{eq:deltathetaevprime}) as 

\begin{align}
    \delta\theta_{E,A} &= \frac{[k(1+z)]^{2n}}{2}\bigg[\frac{\alpha_{2n_0}}{\Lambda_E^{2n}}z_{2n} + \frac{\beta_{2n_0}}{\Lambda_E^{2n-1}}D_{2n+1} + \frac{\mu_{2n_0}}{\Lambda_E^{2n+1}}\int Hdz + \frac{\rho_{2n_0}}{\Lambda_E^{2n+1}}\int (1+z)^{-1}H_zdz\bigg], \label{eq:deltathetaeaint} \\
    \delta\theta_{O,A} &= \frac{[k(1+z)]^{2n+1}}{2}\bigg[\frac{\alpha_{2n+1}}{\Lambda_O^{2n+1}}z_{2n+1} + \frac{\beta_{2n+1}}{\Lambda_O^{2n}}D_{2n+2}+ \frac{\mu_{2n_0 + 1}}{\Lambda_O^{2n+2}}\int Hdz + \frac{\rho_{2n_0+1}}{\Lambda_O^{2n+2}}\int (1+z)^{-1}H_zdz\bigg], \\
    \delta\theta_{O,V} &= \frac{[k(1+z)]^{2m}}{2}\bigg[\frac{\gamma_{2m_0}}{\Lambda_O^{2m}}z_{2m} + \frac{\delta_{2m_0}}{\Lambda_O^{2m-1}}D_{2m+1} + \frac{\nu_{2m_0}}{\Lambda_O^{2m-1}k^2}\int Hdz + \frac{\sigma_{2m_0}}{\Lambda_O^{2m-1}k^2}\int (1+z)^{-1}H_zdz\bigg], \\
    \delta\theta_{E,V} &= \frac{[k(1+z)]^{2m+1}}{2}\bigg[\frac{\gamma_{2m_0+1}}{\Lambda_E^{2m+1}}z_{2m+1} + \frac{\delta_{2m_0+1}}{\Lambda_E^{2m}}D_{2m+2} + \frac{\nu_{2m_0+1}}{\Lambda_E^{2m}k^2}\int Hdz + \frac{\sigma_{2m_0+1}}{\Lambda_E^{2m}k^2}\int (1+z)^{-1}H_zdz\bigg], \label{eq:deltathetaevint}
\end{align}
where the effective distance, $D_{\alpha}$ and the effective redshift parameter $z_{\alpha}$ are defined in Sec.~\ref{sec:genericlagrangian}, and $H_z \equiv dH(z)/dz$, with $H(z) = H_0\sqrt{\Omega_{m,0}(1+z)^3 + \Omega_{r,0}(1+z)^4 + \Omega_{\Lambda,0}}$. 

\noindent
From Eqs.~(\ref{eq:deltathetaeaint})-(\ref{eq:deltathetaevint}), we have

\begin{align}
    h_{\text{R,L}} &= \Bar{h}_{\text{R,L}} \nonumber \\ &\times \text{exp}\bigg\{(\lambda_{\text{R,L}})^{2n}\bigg[\frac{[k(1+z)]^{2n}}{2}\bigg(\frac{\alpha_{2n_0}}{\Lambda_E^{2n}}z_{2n} + \frac{\beta_{2n_0}}{\Lambda_E^{2n-1}}D_{2n+1} + \frac{\mu_{2n_0}}{\Lambda_E^{2n+1}}\int Hdz \nonumber \\ &+ \frac{\rho_{2n_0}}{\Lambda_E^{2n+1}}\int (1+z)^{-1}H_zdz\bigg) + \lambda_{R,L}\frac{[k(1+z)]^{2n+1}}{2}\bigg(\frac{\alpha_{2n_0+1}}{\Lambda_O^{2n+1}}z_{2n+1} + \frac{\beta_{2n_0+1}}{\Lambda_O^{2n}}D_{2n+2} \nonumber \\ &+ \frac{\mu_{2n_0+1}}{\Lambda_O^{2n+2}}\int Hdz + \frac{\rho_{2n_0+1}}{\Lambda_O^{2n+2}}\int(1+z)^{-1}H_zdz\bigg)\bigg]\bigg\} \nonumber \\ &\times \text{exp}\bigg\{i(\lambda_{\text{R,L}})^{2m+1}\bigg[\frac{[k(1+z)^{2m}]}{2}\bigg(\frac{\gamma_{2m_0}}{\Lambda_O^{2m}}z_{2m} + \frac{\delta_{2m_0}}{\Lambda_O^{2m-1}}D_{2m+1} + \frac{\nu_{2m_0}}{\Lambda_O^{2m-1}k^2}\int Hdz \nonumber \\ &+ \frac{\sigma_{2m_0}}{\Lambda_O^{2m-1}k^2}\int (1+z)^{-1}H_zdz\bigg) + \lambda_{\text{R,L}}\frac{[k(1+z)]^{2m+1}}{2}\bigg(\frac{\gamma_{2m_0+1}}{\Lambda_E^{2m+1}}z_{2m+1} + \frac{\delta_{2m_0+1}}{\Lambda_E^{2m}}D_{2m+2} \nonumber \\ &+ \frac{\nu_{2m_0+1}}{\Lambda_E^{2m}k^2}\int Hdz + \frac{\sigma_{2m_0+1}}{\Lambda_E^{2m}k^2}\int (1+z)^{-1}H_zdz\bigg)\bigg]\bigg\}, \label{eq:gw-prop-general}
\end{align}
where again $\Bar{h}_{\text{R,L}}$ is the usual GR expression for the right and left-handed modes. We note here, as we did in Sec.~\ref{sec:genericlagrangian}, that Eqs.~(\ref{eq:deltathetaeaint})-(\ref{eq:gw-prop-general}) are written in terms of the cosmic Hubble parameter, $H = \mathcal{H}/a$.

Furthermore, we can extend the general parameterization in \cite{Daniel:2024lev} to $\mathcal{O}(\mathcal{H}^2)$ for the modified GW velocity. From Eqs.~(\ref{GenParam}) and (\ref{COsCEs}), the GWs satisfy the dispersion relation

\begin{align}
    \omega_{\text{R,L}}^2 = k^2\bigg\{1 &+ (\lambda_{\text{R,L}})^{2m+1}k^{2m-1}\bigg[\frac{\gamma_{2m}}{(\Lambda_Oa)^{2m}}\mathcal{H} + \frac{1}{(\Lambda_Oa)^{2m-1}}\bigg(\delta_{2m} + \frac{\nu_{2m}\mathcal{H}^2 + \sigma_{2m}\mathcal{H}'}{k^2}\bigg)\bigg] \nonumber \\ &+ (\lambda_{\text{R,L}})^{2m+2}k^{2m}\bigg[\frac{\gamma_{2m+1}}{(\Lambda_Ea)^{2m+1}}\mathcal{H} + \frac{1}{(\Lambda_Ea)^{2m}}\bigg(\delta_{2m+1} + \frac{\nu_{2m+1}\mathcal{H}^2 + \sigma_{2m+1}\mathcal{H}'}{k^2}\bigg)\bigg]\bigg\}.\label{eq:moddispgen}
\end{align}
From Eq.~(\ref{eq:moddispgen}), we can find the group and phase velocities of a GW, which are given by $v_g = d\omega/dk$ and $v_p = \omega/k$, respectively. We have
\begin{equation} 
\begin{split} 
    v_g^{\text{R,L}} &= 1 + (\lambda_{\text{R,L}})^{2m+1}k^{2m-1}\bigg\{m\bigg[\frac{\gamma_{2m}}{(\Lambda_Oa)^{2m}}\mathcal{H} + \frac{1}{(\Lambda_Oa)^{2m-1}}\bigg(\delta_{2m} + \frac{\nu_{2m}\mathcal{H}^2 + \sigma_{2m}\mathcal{H}'}{k^2}\bigg)\bigg] \nonumber \\ &+ \lambda_{\text{R,L}}k\bigg(m + \frac{1}{2}\bigg)\bigg[\frac{\gamma_{2m+1}}{(\Lambda_Ea)^{2m+1}}\mathcal{H} + \frac{1}{(\Lambda_Ea)^{2m}}\bigg(\delta_{2m+1} + \frac{\nu_{2m+1}\mathcal{H}^2 + \sigma_{2m+1}\mathcal{H}'}{k^2}\bigg)\bigg]\bigg\} \nonumber \\ &- (\lambda_{\text{R,L}})^{2m+1}k^{2m-1}\bigg\{\frac{\nu_{2m}\mathcal{H}^2 + \sigma_{2m}\mathcal{H}'}{(\Lambda_Oa)^{2m-1}k^2} + \lambda_{\text{R,L}}k\bigg[\frac{\nu_{2m+1}\mathcal{H}^2 + \sigma_{2m+1}\mathcal{H}'}{(\Lambda_Ea)^{2m}k^2}\bigg]\bigg\},
\end{split}
\end{equation}

\begin{equation}
\begin{split}
    v_p^{\text{R,L}} &= 1 + \frac{(\lambda_{\text{R,L}})^{2m+1}k^{2m-1}}{2}\bigg\{\frac{\gamma_{2m}}{(\Lambda_Oa)^{2m}}\mathcal{H} + \frac{1}{(\Lambda_Oa)^{2m-1}}\bigg(\delta_{2m} + \frac{\nu_{2m}\mathcal{H}^2 + \sigma_{2m}\mathcal{H}'}{k^2}\bigg) \nonumber \\ &+ \lambda_{\text{R,L}}k\bigg[\frac{\gamma_{2m+1}}{(\Lambda_Ea)^{2m+1}}\mathcal{H} + \frac{1}{(\Lambda_Ea)^{2m}}\bigg(\delta_{2m+1} + \frac{\nu_{2m+1}\mathcal{H}^2 + \sigma_{2m+1}\mathcal{H}'}{k^2}\bigg)\bigg]\bigg\}.
\end{split}
\end{equation}

\newpage

\bibliographystyle{hunsrt}
\bibliography{bib}

\end{document}